\documentclass[12pt,a4paper]{article}
\usepackage{amsmath,amssymb}
\usepackage{graphicx}
\usepackage[articletitle=true, maxauthors=40]{achemso}

\usepackage{xcolor}
\usepackage{paralist}

\usepackage{mathtools} 
\usepackage{siunitx}
\usepackage{bm}

\begin{document}

\begin{center}

\vspace*{1cm}

{\LARGE\bf Concentration-Flux-Steered Mechanism\\[1ex] Exploration with an Organocatalysis\\[1ex] Application\\[2ex]}

\vspace{.5cm}

{\large Moritz Bensberg and Markus Reiher\footnote{email: markus.reiher@phys.chem.ethz.ch}}\\[2ex]

ETH Z\"urich, Laboratorium f\"ur Physikalische Chemie, Vladimir-Prelog-Weg 2,\\ 8093 Z\"urich, Switzerland\\[4ex]

\vspace*{.51cm}

{\bf Dedicated to Professor Helmut Schwarz on the occasion of this 80th birthday.}
\end{center}

\vspace{.31cm}

\begin{abstract}
    Investigating a reactive chemical system with automated reaction network exploration algorithms provides a more detailed picture of its chemical mechanism than what would be accessible by manual investigation. In general, exploration algorithms cannot uncover reaction networks exhaustively for feasibility reasons. They should therefore decide which part of a network is kinetically relevant under some external conditions given. Here, we
    propose an automated algorithm that identifies and explores kinetically accessible regions of a reaction network on the fly by explicit modeling of concentration fluxes through an (incomplete) reaction network that is emerging during automated first-principles exploration. Key compounds are automatically identified and selected for the continuation of the exploration.
    As an example, we explore the reaction network of the multi-component proline-catalyzed Michael addition of propanal and nitropropene.
    Our algorithm provides a mechanistic picture of the Michael addition in unprecedented detail.  
\end{abstract}

\newpage

\section{Introduction}\label{sec:intro}

Chemical reaction mechanisms are a key to our understanding of chemical reactivity. They represent the essential reaction paths connecting reactants and products in a generally large and complex network of possible chemical reactions. 
For first-principles-based micro-kinetic modeling, the network should be one of elementary reaction steps representing transitions between stationary points on the corresponding Born--Oppenheimer surface. Then, reaction kinetics can be directly characterized by transition state theory\cite{Eyring1935, Truhlar1996}.

Explicit manual investigation of all possible reaction paths is usually unfeasible so that relevant paths may be overlooked. To address the challenge of large reaction networks, a range of automated reaction network exploration algorithms has been presented in the last years (see Refs. \citenum{Sameera2016, Dewyer2017, Vazquez2018, Simm2018, Green2019, Unsleber2020, Maeda2021, Baiardi2021, Steiner2022} for reviews). These algorithms allow for automated explorations of chemical reaction networks with as little manual supervision and intervention as possible.

Even though automated reaction network exploration can, in principle, provide an extensive network of chemical reaction steps, it will, for feasibility reasons, usually not be possible to probe every possible path. Hence, the focus must be on the part of the network that is relevant to the kinetics of the reaction under given external conditions. In this work, we propose a protocol that materializes this focus by short micro-kinetic modeling simulations interleaving a running automated exploration of a reaction network.

Commonly, micro-kinetic modeling is employed as an a-posteriori analysis step after the automated construction of the reaction network has been finished. For instance, the computer programs \textsc{NetGen}\cite{Susnow1997,broadbelt1994computer, Broadbelt1994, Witt2000, Koninckx2022} and \textsc{EXGAS}\cite{Warth2000} automatically construct reaction networks from graph-based reactivity rules, which are then analyzed by micro-kinetic modeling. Similarly, the program \textsc{RMG}~\cite{Gao2016, Goldsmith2017, Liu2021, Johnson2022} generates reaction networks and derives a reaction mechanism by analyzing the concentration fluxes through the network.
Chemically meaningful reaction paths can also be extracted directly from large reaction networks through shortest-path algorithms, including kinetic information on the reaction network\cite{Xie2021, Blau2021, Tuertscher2022}.

Condensing a complicated reaction network into a manageable reaction mechanism is possible with our uncertainty-aware kinetic-modeling program \textsc{KiNetX}\cite{Proppe2016, Proppe2018}. \textsc{KiNetX} propagates uncertainties in the reaction rate constants\cite{Proppe2016} and automatically analyzes the sensitivity of the kinetic model with respect to removing a reaction from it in order to ensure that the final reaction mechanism correctly describes the overall kinetics. 

Steering the automated exploration of a reaction network through thermodynamic data was suggested by Sumiya and Maeda \cite{Sumiya2020}.
Their approach propagates the starting populations with the quasi-steady state approximations and the Boltzmann ratios of each species through the network. However, it is restricted to a single potential energy surface, \emph{i.e.}, the number of atoms in the reaction is fixed during the exploration. In contrast to this previous work, we lift the restriction to a single potential energy surface by explicit micro-kinetic modeling. Furthermore, it can be expected that this explicit kinetic modeling should provide a more rigorous picture of the kinetics.

Analyzing micro-kinetic modeling simulations to steer the generation of a reaction network was proposed before by Susnow \emph{et al.}\cite{Susnow1997}
in 1997 and employed in the context of combustion chemistry\cite{Matheu2001, Geem2006, Blurock2012}..
In this earlier work, a reaction mechanism was constructed in an iterative way by successively including all compounds into the mechanism that show a significant formation rate during micro-kinetic modeling simulations on the incomplete reaction network during exploration. Here, we extend this earlier work and exploit additional information from the micro-kinetic modeling.
We exploit the maximum concentrations and the magnitude of the concentration flux passing through the compounds. Analyzing the maximum concentrations allows us to identify compounds that are highly populated at some point during the micro-kinetic modeling, making bimolecular reactions of these compounds probable because they are likely to meet other compounds. By contrast, a high concentration flux for a compound signals that the compound is easily accessible but must not necessarily have a long lifetime. For these compounds, unimolecular reactions should be probed.
In addition to this extended analysis of the micro-kinetic modeling simulations, we do not rely on heuristic reaction rate rules common for combustion chemistry\cite{Susnow1997, Matheu2001, Geem2006}
but obtain all reaction rate constants from first-principles calculations (in an automated fashion).

This work is organized as follows: in the next section, we introduce the organocatalysis example, which serves the purpose of demonstrating the capabilities of our kinetics-interlaced exploration algorithm (KIEA). It was also selected to provide in-depth mechanistic information for this important reactive system.
Then, we describe our concentration-flux-steered exploration ansatz in sections~\ref{sec:ansatz}--\ref{sec:heuristic_rules}. How we refined important reaction steps is highlighted in section~\ref{sec:double_ended_refinement}. The explored network and reaction mechanism is then presented in section~\ref{sec:results}.

\section{Kinetics-interlaced Exploration Algorithm}

\subsection{Development-driving Example: Multi-component Organo\-catalysis}\label{organo}

The example at which we developed KIEA is the multi-component, proline-catalyzed Michael addition of propanal and nitropropene, which is a prime example of asymmetric organocatalysis.
Since the pioneering work on asymmetric proline-catalyzed aldol reactions by List and Barbas\cite{List2000}, asymmetric organocatalysis\cite{Mukherjee2007, Seayad2005} has seen a swift development that was even compared to the American ``gold rush''\cite{Melchiorre2008}. Since the organic catalysts are generally non-toxic, inexpensive, stable, and metal-free, they are especially attractive for chemical synthesis. 

\begin{figure}[!ht]
    \centering
    \includegraphics[width = 0.5\textwidth]{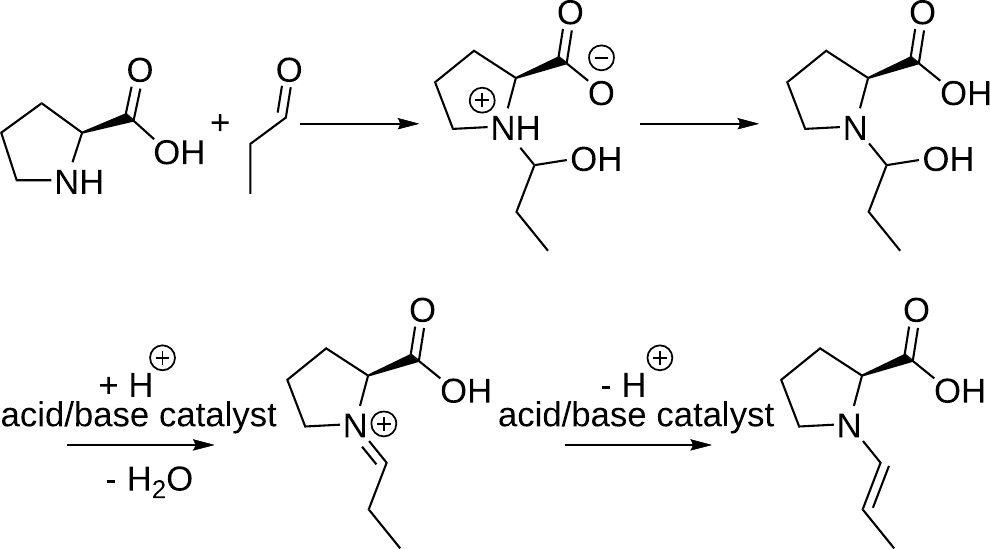}
    \caption{Enamine formation from proline and propanal.}
    \label{fig:enamine_formation}
\end{figure}

\begin{figure}
    \centering
    \includegraphics[width = 0.9\textwidth]{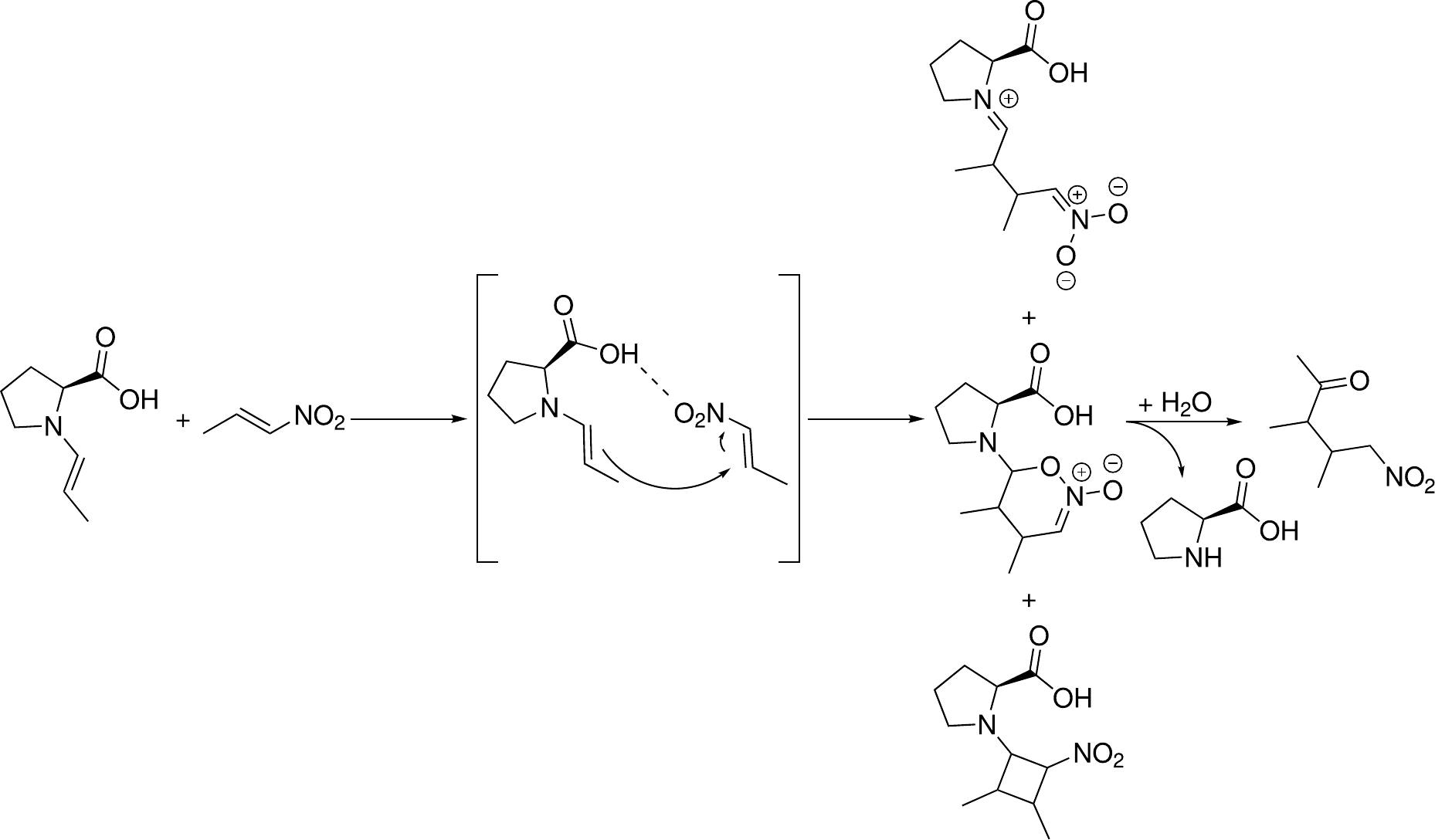}
    \caption{Michael addition of the intermediate enamine with nitropropene.}
    \label{fig:michael_addition}
\end{figure}

The proline-catalyzed Michael addition\cite{List2001} of aldehydes to nitro-olefines has been investigated in previous studies\cite{Maillard2019, CastroAlvarez2019, Sahoo2012, Seebach2013, Vilarrasa2017}. In the mechanism proposed for both reactions, catalytic active enamine species are formed by nucleophilic attack of proline's amine group at the carbonyl carbon atom, followed by water elimination and subsequent enamine formation, as shown in Fig.~\ref{fig:enamine_formation}. The carboxyl group of the catalytically active enamine induces the desired stereochemistry by functioning as a directing group for the subsequent attack of the second carbonyl or olefine.

The reaction mechanism of the proline or prolinol ether-catalyzed Michael addition of aldehydes to nitro-olefines then proceeds with the enamine electrophilically attacking the olefine, as shown in Fig.~\ref{fig:michael_addition}, which leads to the formation of intermediate zwitterionic species \textbf{zw}, intermediate cyclobutanes \textbf{cb}, and intermediate dihydrooxazine-N-oxides \textbf{dx}\cite{CastroAlvarez2019, PatoraKomisarska2011, Bures2016, Seebach2013}. These intermediates are then hydrolyzed, releasing the catalyst and forming the nitro-alkene. For the hydrolysis of all three intermediates, reaction mechanisms have already been reported (zwitterion\cite{Seebach2012}, cyclobutane\cite{Bures2016}, dihydrooxazine-N-oxides\cite{Gurubrahamam2016}), making the explicit reaction path solvent and reactant specific.

These three different intermediates directly demonstrate how many different reaction paths can become relevant when investigating the reaction mechanism. For instance, three different hydrolysis paths were discussed in Ref.~\citenum{Seebach2013} for the zwitterionic intermediate \textbf{zw} and one for the dihydrooxazine-N-oxides \textbf{dx}. By contrast, the cyclobutane \textbf{cb} intermediate was considered a nonreactive  species that first must undergo ring-opening to the zwitterionic intermediate \textbf{zw} before hydrolysis. Later, Bures \emph{et al.}\cite{Bures2016} argued that the cyclobutane species \textbf{cb} may be hydrolyzed directly and functions as a stable intermediate of which only one stereoisomer is reactive, hence controlling the enantioselectivity of the reaction.

\subsection{Exploration Protocol}\label{sec:ansatz}

In this work, we establish a detailed reaction network of the proline-catalyzed Michael addition of nitropropene and propanal based on the automated reaction network exploration algorithms implemented in \textsc{SCINE Chemoton}\cite{Unsleber2022, Bensberg2022b}.  
We start the exploration with the initial reactant species only, probe for possible reactions, obtain new compounds, and continue the exploration from these newly explored compounds. To avoid exploring parts of the reaction network that are irrelevant under reaction conditions, we calculate the concentration flux through the (incomplete) reaction network by short micro-kinetic modeling simulations. Based on the concentration data obtained from the kinetic modeling simulations, our approach automatically selects kinetically easily accessible compounds for further exploration. In this way, we automatically explore the reaction network in a target-oriented way.

We encode the reaction network according to the concepts and terminology discussed in Refs.~\citenum{Simm2017}, \citenum{Unsleber2020}, and \citenum{Unsleber2022}: \emph{structures} arrange into \emph{compounds}, \emph{elementary steps} are combined to \emph{reactions}, \emph{structures} are prepared for reactions in \emph{flasks}. Structures are stationary points on the Born-Oppenheimer potential energy surface (PES) with a fixed set of atoms, charge, and multiplicity. A compound is the collection of all structures with the same charge, multiplicity, and molecular bond/abstract graph representation as defined in Ref.~\citenum{Sobez2020} that constitute a single molecule. 
Elementary steps represent the reactive process of transforming between two sets of structures, generally as the transition from one valley on the PES to another valley through a single transition state. 
However, such a step may also be barrier-less, \emph{e.g.}, for the formation of a weakly interacting complex from two isolated molecules. Hence, \textsc{Chemoton} considers the formation of reactive pre-complexes as elementary steps. 
Reactions collect all elementary steps that transform between structures of the same compounds or flasks. For instance, two elementary steps that connect different possible minimum structures (conformers) of proline to different structures of proline's zwitterionic form (deprotonated carboxyl group and protonated amine group) are assigned to the same reaction because, in both cases, the compounds on the left-hand side and the compounds on the right-hand side of the elementary steps are the same. To avoid any confusion about the notation, we note that in the micro-kinetic modeling literature(\emph{e.g.} see Ref.~\citenum{Green2019}) our definition of a compound corresponds to a \emph{species} and structures are referred to as \emph{conformers}.

Our exploration strategy consists of two distinct steps: (i) reactant selection and (ii) autonomous exploration. In the reactant selection step, a set of compounds is selected and assigned a starting concentration. In the autonomous exploration step, unimolecular and bimolecular reactions are probed for all reactant compounds as described in Ref.~\citenum{Unsleber2022}. After this initial set of reaction trials, a micro-kinetic modeling simulation, (iii), is carried out for the already established (though incomplete) network. Then, all compound pairs $nm$ will be considered for further bimolecular reaction trials if they have a high probability for an encounter. An upper bound for the probability is given by the product of their maximum concentrations, $c_\mathrm{max}^{n}$ and $c_\mathrm{max}^{m}$, observed in the micro-kinetic modeling simulation (see below). Hence, the condition for bimolecular reaction trials reads $c_\mathrm{max}^{n} c_\mathrm{max}^{m} > \tau_\mathrm{max}$, where $\tau_\mathrm{max}$ is a given threshold. 
Unimolecular reactions will be probed if the compound is visited frequently during kinetic modeling, \emph{i.e.}, if the integrated flux through the compound is larger than a given threshold ($c_\mathrm{flux}^{n} > \tau_\mathrm{flux}$). 
The interlaced cycles of reaction trials (i)+(ii) and micro-kinetic modeling (iii) of KIEA are continued until no further reaction trials need to be considered according to the criteria (bimolecular reactions: $c_\mathrm{max}^{n} c_\mathrm{max}^{m} > \tau_\mathrm{max}$, unimolecular reactions: $c_\mathrm{flux}^{n} > \tau_\mathrm{flux}$) stated above, 
and the exploration is considered finished in this step. Then, the next reactant selection step is performed, and the autonomous exploration is restarted.

To increase the efficiency of the exploration for our specific example, we restricted the process to chemically meaningful reaction paths 
and exploited chemical intuition (see below for details). 
For this purpose, we decided on a step-wise algorithm that couples the automated exploration with manual intervention and selection of compounds. This option allows for greater
flexibility as the whole range between full autonomy and interactive manual steering is accessible. Furthermore, the option allowed us to efficiently address reactions in which a crucial exergonic product was hidden behind an intermediate that 
was comparatively high in energy. While this intermediate may have an insufficient concentration flux to be selected by the automated exploration algorithm, it can be identified 
during the compound selection step and selected as the starting point for a new automated exploration run.

Next, to maximize the efficiency of the exploration, we exploited a hierarchy of electronic structure methods. For the exploration, we calculated structures with the computationally efficient  GFN2-xTB\cite{Bannwarth2019} method, which is a tight-binding model
of limited accuracy for structures and energies (in the context of organic chemistry especially for transition states). Especially the former, i.e, inaccurate structures, are worrisome in an exploration protocol that relies on structural data to identify stationary points on a PES (improved tight-binding models or emerging universal machine learning potentials will be a way out in the future). 
However, we found for our example that the GFN2-xTB structures were reasonable and sufficiently reliable for further refinement (no odd structures with unusual bonding
patterns were found that can be a consequence of the
approximation inherent to the tight-binding electronic 
structure model GFN2-xTB).
The micro-kinetic modeling we then based on electronic energies calculated for these structures with density-functional theory (DFT). We avoided any structure reoptimization with DFT or calculation of free energy corrections during the exploration to keep the computational cost of the exploration as low as possible. 

The accuracy of our data was then increased in a second refinement step after completing the network exploration to provide a quantitatively meaningful description of the reaction mechanism.
We selected a reaction path guided by the concentration flux through the network from the final reaction network and refined this path with a double-ended reaction exploration strategy, including reoptimization of the structures and recalculation of the intrinsic reaction coordinate with DFT and single point energies from domain-based local pair natural orbital coupled cluster with singles, doubles, and perturbative triples excitations [DLPNO-CCSD(T)]\cite{Riplinger2016, Riplinger2013a} for die stationary points on the reaction path.

\subsection{Micro-kinetic Modeling\label{sec:micro-kinetic_modeling}}

The purpose of the micro-kinetic modeling during the reaction network exploration is the prediction of which compounds are likely to be relevant for further exploration. 
Compared to purely thermodynamics-based approaches\cite{Sumiya2020, Tuertscher2022} the micro-kinetic 
modeling provides a time resolution of the concentration fluxes identifying not only which compounds are accessible, but also which compounds show significantly high concentrations during the reaction and should be probed for bimolecular reactions.
For feasibility reasons, we need to employ a simplified, computationally efficient definition of the reaction rate constants. The forward ('$+$') and backward ('$-$') rate constants $k_I^{+/-}$ for a reaction $I$ are defined according to Eyring's absolute rate theory\cite{Eyring1935, Truhlar1996} as
\begin{align}
    k^{+/-}_I = \Gamma \frac{k_B T}{h} \exp\left[- \frac{\Delta G_I^{\ddagger +/-}}{k_B T}\right]~,
\end{align}
where Eyring's fudge factor $\Gamma$ is the so-called transmission coefficient, $\Delta G_I^{\ddagger +/-}$ are the free energies of activation for the two directions of $I$, $k_B$ is Boltzmann's constant, and $T$ the temperature.
To make our model as simple as possible, we set $\Gamma=1$ and approximate $\Delta G_I^{\ddagger +/-}$ by the electronic energy of activation $\Delta E_I^{\ddagger +/-}$ (including the contribution from the implicit solvent model) only
\begin{align}
    k_I^{+/-} \approx \frac{k_B T}{h}\exp\left[- \frac{\Delta E_I^{\ddagger +/-}}{k_B T}\right]~.
\end{align}
We assume that barrier-less reactions and reactions with a very low barrier are diffusion controlled. These are described by constant maximum rate constants, which are taken to be larger for unimolecular than for bimolecular reactions (see Sec.~\ref{sec:comp_details}).
In general, the description of the degrees of freedom in the condensed phase by the rigid-rotor, particle in a box, and harmonic oscillator approximations is conceptually inappropriate because it this standard model cannot properly account for the restrictions on the motion imposed by the solvent cage\cite{Reiher2019}.
Note that bimolecular reactions are described in \textsc{Chemoton} by an initial barrier-less elementary step forming a reactive complex followed by an elementary step crossing the reaction barrier. A reactive complex would be considered if it was stabilized by at least $12.0~\si{kJ.mol^{-1}}$ in terms of its electronic energy compared to the separated reactants.

\begin{figure}
    \centering
    \includegraphics[width=0.9\textwidth]{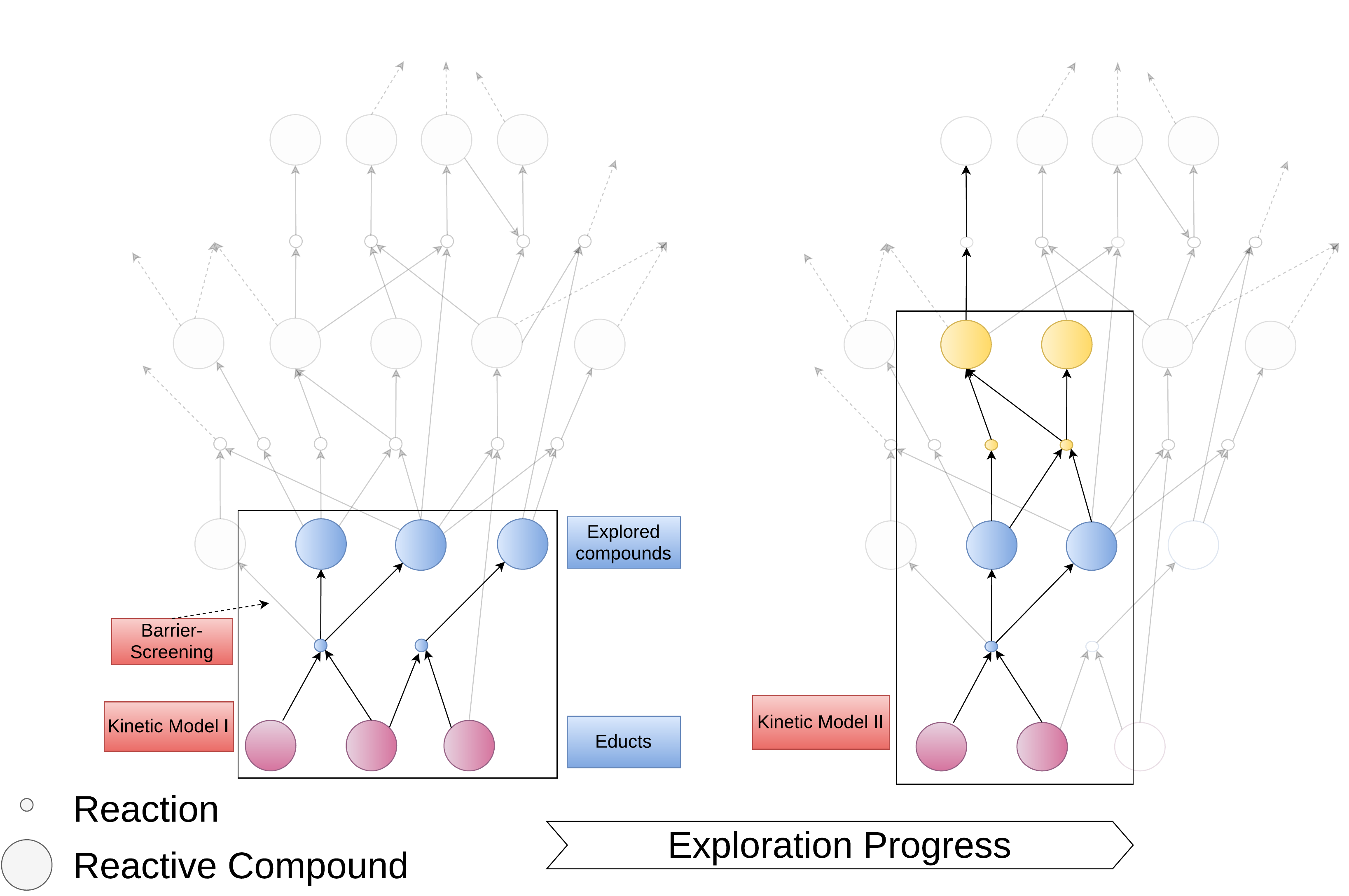}
    \caption{Kinetic model restriction during exploration. Reactions with a large barrier are omitted from micro-kinetic modeling, as are reactions that had a negligible reaction flux in the micro-kinetic modeling of a previous exploration step.}
    \label{fig:kinetic_modeling_trimming}
\end{figure}

In the micro-kinetic modeling, the system of ordinary differential equations encoded by the reaction network is integrated numerically with the Runge-Cutta-type algorithm proposed in Ref.~\citenum{Niemeyer2014}. This scheme includes an automated integration step selection. The intermediate quantities, 
such as concentration gradients, required for the integration provide additional insights into the kinetics of the reaction network and form the basis for the integrated flux $c_\mathrm{flux}^{n}$ exploited for compound selection in the automated reaction network exploration. Concentration gradients and concentration fluxes 
are defined in analogy to Ref.~\citenum{Proppe2018}.
The forward/backward reaction rate $f_I^{+/-}(t)$ at some time $t$ is given with the concentrations $c_n(t)$ and the stoichiometric coefficients $S^{+/-}_{nI}$ for the species $n$ and reaction $I$ by
\begin{align}
    f_I^{+/-}(t) = k_I^{+/-} \prod_{n=1}^N [c_n (t)]^{S^{+/-}_{nI}}~.
\end{align}
Hence, the net concentration transport $f_I(t)$ of reaction $I$ is given as
\begin{align}
    f_I(t) = f_I^{+}(t) - f_I^{-}(t)~,
\end{align}
the net stoichiometry of the reaction as
\begin{align}
    S_{nI} = S^{-}_{nI} - S^{+}_{nI}~,
\end{align}
and the concentration gradient $g_n(t)$ for each species $n$ at time $t$ by
\begin{align}
    g_n(t) = \sum_I S_{nI} f_I(t)~.
\end{align}
The concentration gradient is integrated numerically to yield the concentrations $c_n(t)$
\begin{align}
    c_n(t_\mathrm{max}) = c(t_0) + \int_{t_0}^{t_\mathrm{max}} \mathrm{d}t~ g_n(t)~,
\end{align}
where $t_0$ and $t_\mathrm{max}$ are the start and end times, respectively.

These definitions allow us to define the total concentration flux $c_\mathrm{flux}^n$ for any species and the total reaction edge flux $F_I$ for any reaction from the intermediate $f_I(t)$ as
\begin{align}
    F_I = \int_{t_0}^{t_\mathrm{max}} \mathrm{d}t~\left| f_I(t) \right|
    \label{eq:reaction_edge_flux}
\end{align}
and
\begin{align}
    c_\mathrm{flux}^n = \sum_I \left(S^{-}_{nI} + S^{+}_{nI}\right) F_I~.
\end{align}

Many reactions will be discovered during an automated exploration of a reaction network, which we cannot all include in the kinetic model. Therefore, we pruned the model as shown in Fig.~\ref{fig:kinetic_modeling_trimming}. We required that only paths with an activation barrier lower than a maximum value $\Delta E_\mathrm{max}^{\ddagger}$ are traversed when starting from the initial reactant compounds. Furthermore, we performed many consecutive kinetic model simulations during the exploration. Therefore, we could analyze the concentration fluxes $c_\mathrm{flux}^n$ to detect reactions that had only a negligible flux in the previous run to omit them in the next simulation. In this way, we identified reactions with a relatively low barrier, which were, however, significantly slower than competing reactions and, therefore, not key to the chemical kinetics through the network.
Similar to the kinetic model reduction ansatz discussed in Ref.~\citenum{Proppe2018}, we had omitted reactions from the micro-kinetic modeling if any species in the reaction showed a concentration smaller than some threshold, $c_\mathrm{flux}^n < \tau_\mathrm{flux}^\mathrm{kin}$.

\subsection{Heuristic Reaction-trial Restriction\label{sec:heuristic_rules}}

Bimolecular and unimolecular reactions with up to 60 atoms were probed with the algorithm described in Ref.~\citenum{Unsleber2022}. To restrict the reaction trials in these searches for elementary steps to a chemically meaningful selection, we considered two-bond modifications in the reaction trials. Of these modifications, at most one was allowed to be a dissociation reaction and two to be association reactions. We primarily focussed on already known reactivity patterns and then defined atoms that were considered reactive through a set of rules. 
These rules allowed us to investigate the typical protonation, carbonyl, and enamine chemistry that one would expect based on organic textbook knowledge while being general enough to allow for unexpected reactivity. Note that these rules only restricted which reaction coordinates were probed but not which reactions were discovered by the automated algorithm.
Furthermore, we explicitly allowed reaction trials for the direct deprotonation of intermediate \textbf{cb} and the ring-opening of intermediate \textbf{dx} (see Fig.~\ref{fig:michael_addition}).

\begin{figure}
    \centering
    \includegraphics[width=0.8\textwidth]{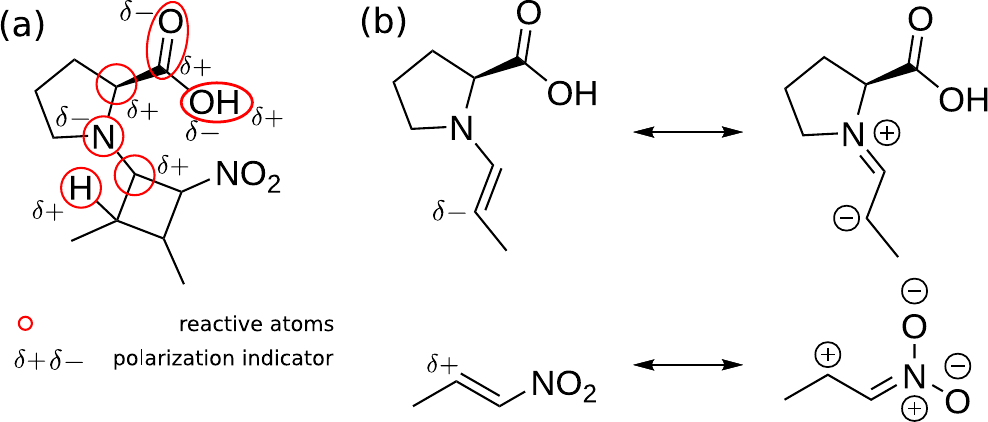}
    \caption{Illustration of the heuristic reaction rules applied in this work. (a) The highlighted atoms and functional groups are considered reactive according to the rules given in Sec.~\ref{sec:heuristic_rules}. (b) Illustration of the assignment of the polarization $\delta +$ and $\delta -$ for the enamine (top) and nitropropene (bottom) according to Pauling electronegativities.}
    \label{fig:reaction_rules}
\end{figure}

Note that similar heuristic restrictions have been used in graph-based exploration of reaction networks reported in Refs.~\citenum{Ismail2019} and \citenum{Rappoport2014}. In general, heuristic reaction rules can also be derived directly from the electronic structure of the reactants to provide generally applicable ’first-principles heuristics’ \cite{Bergeler2015} for the exploration process. Reactivity concepts derived from electronic structure can be derived from partial charges, Fukui functions, the electrostatic potential and other chemical concepts \cite{Bergeler2015, Grimmel2019, Grimmel2021}. However, we will restrict our approach here to the heuristic rules detailed below for the sake of simplicity, because these selection rules are neither key to the development of our concentration-flux-steered exploration algorithm, nor will they limit the results found for our organocatalysis example (since the rules apply well to this type of organic chemistry):

Hydrogen: Hydrogen atoms would be considered for reaction trials if they were in a typical acidic position. For this, we selected hydrogen atoms in the 2-position to either carbonyl groups (excluding carboxyl groups), imine groups, or acetal groups, as well as the acidic hydrogen atoms in carboxyl groups, 
protonated amine groups,
and the hydrogen atoms in the cyclobutane-like intermediates, as shown in Fig.~\ref{fig:reaction_rules}(a).

Carbon: Carbon atoms would be considered for reaction trials if they were part of carbonyl or imine groups, bonded to two oxygen or nitrogen atoms, or if they were next neighbors of a carbon atom selected as reactive according to these first rules. 
Furthermore, olefinic carbon atoms and the carbon atom highlighted for the cyclobutane-like intermediate in Fig.~\ref{fig:reaction_rules}(a) were considered as reactive in elementary step searches (i.e., in the reaction trials).

Oxygen: Oxygen atoms would be considered reactive if they were part of a hydroxy or carbonyl group. Furthermore, oxygen atoms in the cycle of dihydrooxazine-N-oxide-like intermediates were considered reactive.

Nitrogen: Only nitrogen atoms in amine groups were considered reactive.

To limit the number of reaction trials and further accelerate the exploration, we allowed only reactions between atoms that are
differently polarized. Therefore, we assigned polarization indicators, $\delta +$ or $\delta -$, to the atoms. These indicators were determined by the difference in the Pauling electronegativities of the specific atom and its neighbors. If the Pauling electronegativity difference for an atom-pair exceeded an absolute value of $0.4$, the polarization would be assigned this signed difference. A minimum absolute difference was selected to prevent polarization assignment to CH$_x~(x=1, 2, 3)$ groups. Moreover, heuristic polarization rules were introduced to include electronic polarization effects that were not captured by simple electronegativity differences.
These rules are illustrated in Fig.~\ref{fig:reaction_rules}. Note that such polarization indicators could, in principle, be derived from calculated atom-wise partial charges. However, there is no guarantee that such partial charges reflect the reaction rules introduced above. Since the number of rules is very limited, we manually assigned polarization indicators to these reactive-atom rules.

All hydrogen atoms, which were considered reactive according to the reactive-atom rules stated above, were assigned a positive polarization indicator $\delta +$. 
Reactive carbon atoms were assigned a $\delta +$, excluding the olefinic carbon atoms. To model mesomeric effects, the olefinic carbon
atoms would be assigned a $\delta -$ if they were either in a 2-position to an amine or in a 3-position to a nitro group. Amine groups were considered electron-rich and their nitrogen atoms were assigned a negative polarization indicator $\delta -$.

\subsection{Double-ended Elementary Step Refinement\label{sec:double_ended_refinement}}

\begin{figure}
    \centering
    \includegraphics[width=\textwidth]{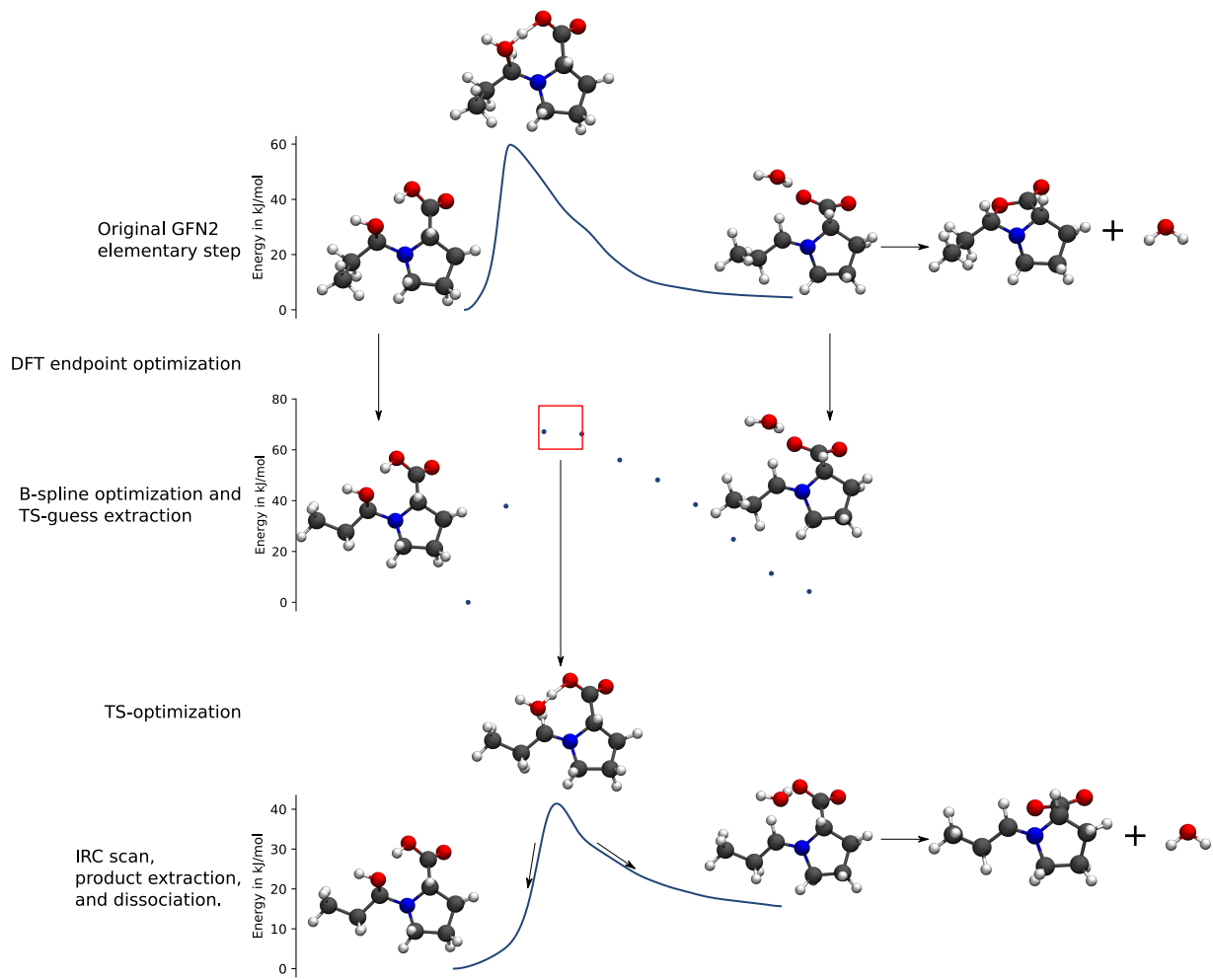}
    \caption{Double-ended structure refinement approach:
             the end points of the original elementary step are reoptimized. Between these end points, a new reaction
             path is optimized and a transition state guess is extracted. This guess is then optimized and the IRC
             is calculated to obtain the products and reactants corresponding to the transition state.}
    \label{fig:struc_refinement}
\end{figure}

Our double-ended elementary step refinement strategy is illustrated in Fig.~\ref{fig:struc_refinement}. The reactants and products of the original elementary step were extracted as the endpoints of the intrinsic reaction coordinate (IRC) scan of the original elementary step found with the initial exploration method, in this case with GFN2-xTB\cite{Bannwarth2019}. 
These end points were then reoptimized with DFT (see Sec.~\ref{sec:comp_details} for details), 
and an initial guess for its reaction path is obtained from linear interpolation between them. The reaction path was then optimized by curve optimization, a specific path optimization scheme introduced in Ref.~\citenum{Vaucher2018}.
Then, a transition state guess was extracted as the maximum along this path. The transition state was optimized from this guess and the IRC 
was calculated to identify the end points (reactants and products) corresponding to a specific transition state. We then calculated the bond orders and partial charges of the end points, constructed an idealized molecular graph from the bond orders, and identified non-bonded molecular fragments.

As an example reactive system, we show the refinement of the intermolecular protonation of the hydroxy group by the carboxyl group (equivalent to the third step in Fig.~\ref{fig:enamine_formation}) of the initial proline propanal addition product in Fig.~\ref{fig:struc_refinement}. Because the hydroxy group is protonated, water dissociates from the weakly interacting complex on the right-hand side of the first elementary reaction step, illustrated by the bottom spline interpolation in Fig.~\ref{fig:struc_refinement}.
To describe the dissociation of water from this complex after the refinement, the molecular fragments were then optimized separately.
For this example, two new elementary steps were added to the database. The first elementary step described the intramolecular protonation of the OH group to form water and the zwitterionic imine through the transition state shown in Fig.~\ref{fig:struc_refinement}. 
The second elementary step shown in Fig.~\ref{fig:struc_refinement}, describes
the barrier-less dissociation of the product complex to form separated water and the imine. Note that the products of this second elementary step differ from the corresponding elementary step optimized with the initial exploration method. Originally, the ring closure of the imine to yield the oxazolidinone through addition of the COO$^-$ group to the imine-carbon atom was favored. The initial zwitterionic structure is not a local minimum on the non-refined potential energy surface but only on the refined surface. Here, the ring closing is not observed.

Furthermore, two special cases were considered in the reaction
refinement:
(A) Optimizing the initial end points led to the same molecule or complex. If at least one of the original endpoints corresponded to separated molecules, a barrier-less elementary step between the separated and reoptimized fragments and the optimized end point species would be added to the database.
(B) If any of the original end points was a complex that corresponded to a different flask after its optimization (e.g., a protonation step occurs barrier-less during the optimization), a barrier-less elementary step between the separated and optimized fragments of the original end point and the optimized endpoint structure would be added to the database.

These two special cases were considered barrier-less processes during the refinement since the original structures, obtained with the fast, but approximate tight-binding approach, might not have corresponded to minimum structures with the refinement method, \emph{i.e.}, DFT. Therefore, the products and reactants found by the refinement may describe slightly different reactions than the initial reactions. Hence, the refinement may result in reactions and compounds impossible to identify with the initial exploration method. However, this is a design feature as it shows how the more accurate, but computationally more demanding approach can correct for deficiencies in the fast, but less accurate model.

\section{Computational Details\label{sec:comp_details}}
All calculations were steered or carrioed out with modules from our open-source \textsc{Scine} software suite.
The reaction trial calculations were set up by \textsc{Scine}-\textsc{Chemoton}\cite{Unsleber2022, Bensberg2022b} and calculated with the \textsc{Scine}-\textsc{Puffin}\cite{Bensberg2022d} and \textsc{Scine}-\textsc{ReaDuct}\cite{Brunken2022, Vaucher2018} programs. The electronic structure and nuclear gradients for the reaction trials were calculated with GFN2-xTB\cite{Bannwarth2019}
combined with the generalized Born and surface area (GBSA)\cite{Onufriev2004, Sigalov2006} implicit solvation model, explicitly parameterized for GFN2-xTB\cite{Bannwarth2019}, 
to model acetonitrile. 

All reaction trial calculations that failed because of an unsuccessful transition state optimization were restarted with acetone as the solvent in GBSA to increase the number of successfully discovered elementary reaction steps by this perturbation of the PES.
Acetone was selected as the second solvent because acetonitrile and acetone are both non-protic solvents with dielectric constants that are different enough to perturb the PES (dielectric constant of acetone is $20.7$ and that of acetonitrile is $37.5$).

It is important to emphasize that the structures of the reactions discussed in the final mechanism (Sec.~\ref{sec:proline_catalyst_michael_addition}) were reoptimized after the exploration during the double-ended step refinement and all energies that enter the micro-kinetic modeling during the exploration were calculated with DFT in a first refinement step.

As this first refinement step, the electronic energies of the stationary points of the reactions discovered with GFN2-xTB(acetonitrile/acetone) would be calculated with DFT if the reaction barrier was at most $100~\si{kJ.mol^{-1}}$.
The electronic energies were calculated with the quantum chemistry program \textsc{Turbomole}\cite{Ahlrichs1989,turbomole741}, the exchange--correlation functional of Perdew, Burke, and Ernzerhof, known as PBE\cite{Perdew96}, with Grimme's D3 dispersion correction\cite{Grimme2010} and Becke--Johnson damping\cite{Grimme2011}, the def2-SVP basis set\cite{Ahlrich2005}, and the conductor-like screening model (COSMO)\cite{Klamt1993, Klamt2011} to model acetonitrile (with a dielectric constant of $37.5$) as the solvent.

The reaction fluxes were modeled with the C++-reimplementation of the program \textsc{Scine}-\textsc{KiNetX}\cite{Proppe2018, Proppe2022} based on the DFT electronic energies. 
This version of \textsc{KiNetX} integrates the concentrations with the Runge-Kutta approach proposed in Ref.~\citenum{Niemeyer2014} with an automated time step selection. Each of the short kinetic modeling simulations was integrated for a total of $5\cdot 10^{7}$ time steps. Reactions would be excluded from the kinetic modeling, as discussed in Sec.~\ref{sec:micro-kinetic_modeling}, if they featured a barrier of more than $\Delta E_\mathrm{max}^{\ddagger} = 80.0~\si{kJ.mol^{-1}}$ or a concentration flux of less than $\tau_\mathrm{flux}^\mathrm{kin} = 1 \cdot 10^{-5}$ in a previous kinetic modeling run.  Barrier-less reactions were assigned a constant reaction rate corresponding to a barrier of $4.18~\si{kJ.mol^{-1}}$ for unimolecular reactions and $12.0~\si{kJ.mol^{-1}}$ for bimolecular reactions, as discussed in Sec.~\ref{sec:micro-kinetic_modeling}. Note that we only aim at an qualitative picture of the reaction kinetics which we can calculate efficiently. For this reason, we consider the electronic energy barriers only, as well as barrier-less reactions.

For the initial step of the exploration that produced the enamine (see Fig.~\ref{fig:enamine_formation}), the thresholds for selecting compounds as reactive were chosen as $\tau_\mathrm{max} = 1\cdot 10^{-2}~\si{mol.L^{-1}}$ and $\tau_\mathrm{flux} = 1\cdot 10^{-3}~\si{mol.L^{-1}}$. To reduce the number of reaction trials and the computational cost, the network was explored in a more narrow and target-oriented way by increasing the thresholds to $\tau_\mathrm{max} = 1\cdot 10^{-1}~\si{mol.L^{-1}}$ and $\tau_\mathrm{flux} = 1\cdot 10^{-2}~\si{mol.L^{-1}}$, in the following steps. The main motivation for choosing $\tau_\mathrm{flux}$ and $\tau_\mathrm{max}$ is to limit the number of reaction trials to a number that is feasible to probe within one day between micro-kinetic modeling steps with a fast quantum chemical approach such as GFN2-xTB. A detailed study of the dependence of these parameters on the exploration of a specific system we defer to future work.

The starting concentrations for the first exploration step (step I, enamine formation) were chosen as $1.0~\si{mol.L^{-1}}$ for proline and for
propanal. In the second exploration step (step II, Michael addition), the starting concentrations were set to $0.5~\si{mol.L^{-1}}$ for both cis- and trans-enamine isomers and to $1.0~\si{mol.L^{-1}}$ for nitro propene. In the third exploration step (step III, intermediate water addition), the concentration for intermediate \textbf{dx} (see Fig.~\ref{fig:michael_addition} and water were set to
$1.0~\si{mol.L^{-1}}$ and for proline to $0.5~\si{mol.L^{-1}}$. Proline was included to facilitate Br{\o}nsted acid/base chemistry during the exploration.
In the last exploration step (step IV, catalyst recovery), the concentration for proline was kept at $0.5~\si{mol.L^{-1}}$, and the concentration of the water adduct (see Fig.~\ref{fig:full_energy_landscape}, compound \textbf{w-m}) was set to $1.0~\si{mol.L^{-1}}$. Note that these concentrations do not reflect the experimental conditions in which nitropropene and propanal are present in large excess compared to proline. If we set experimental concentrations at the start of the exploration, the automated exploration would have unfolded the chemistry of propanal, because of the tiny amount of the proline catalyst, which would, at the beginning of the exploration, not have been identified as a catalyst (this is only possible after the catalytic cycle has been discovered). Therefore, we deliberately chose the concentrations for the compounds to be similar in order to facilitate the exploration of the catalytic cycle, for which we did not expect nitropropene or propanal to be required in significant excess; that is, we expected that at most one or two propanal/nitropropene molecules take part explicitly in the catalytic cycle.
Furthermore, we did not consider the role of typical additives\cite{Pihko2004, Pihko2006, Zotova2007} , such as water, during the exploration with exception of step III, because the reaction will proceed even if no water is added (though at reduced yield).

\begin{figure}
    \centering
    \includegraphics[width=\textwidth]{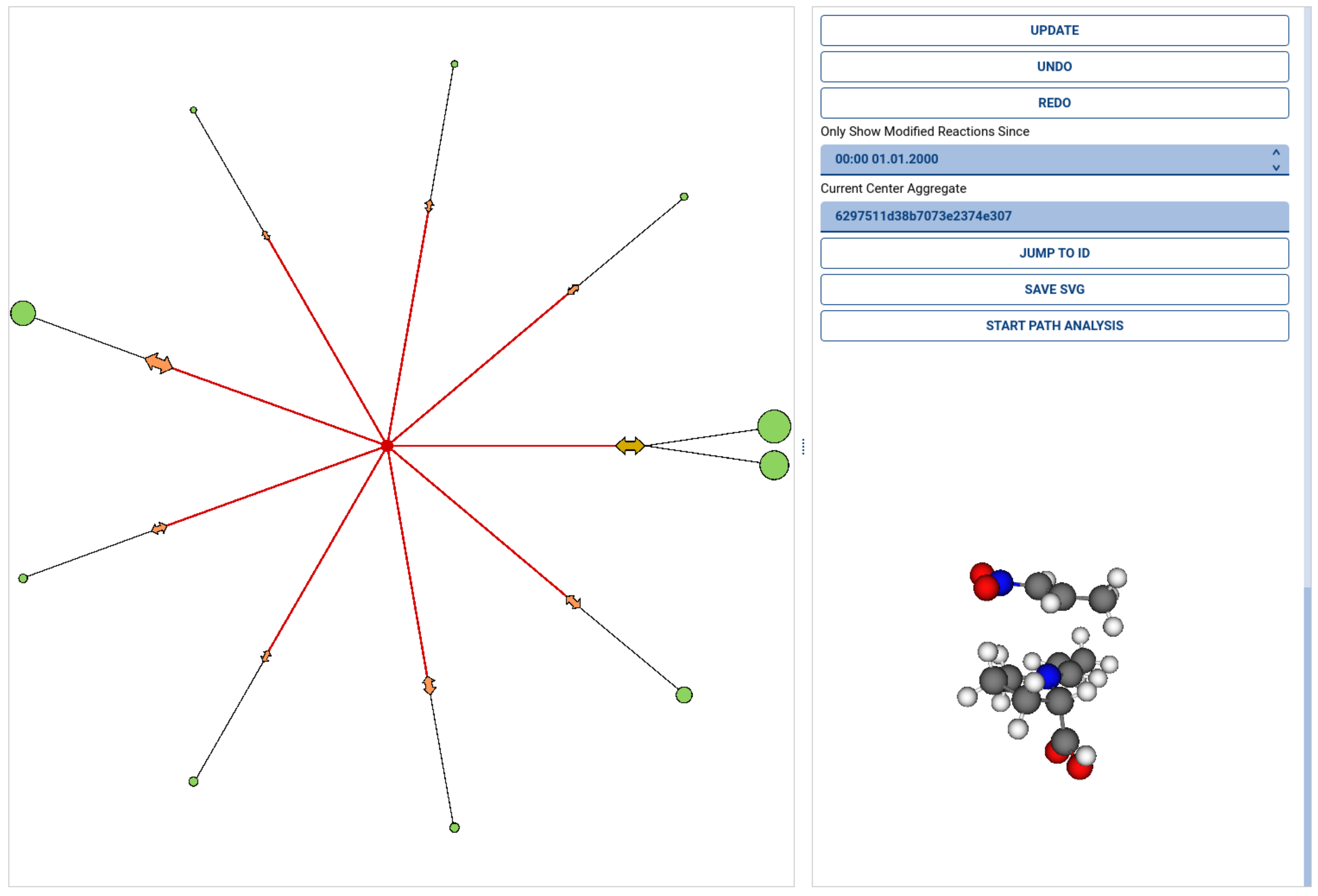}
    \caption{The interactive visualization of the reaction network in the graphical user interface \textsc{Heron}. The orange arrows represent reactions with barriers, the yellow arrows represent barrier-less reactions, and the green circles represent compounds. The central circle was manually selected and the molecular structure in this flask is shown on the right-hand side. The circle sizes indicate the maximum concentration of the corresponding compound. The arrow sizes indicate the concentration flux $F_I$. Double-clicking the circle moves it into focus, making it the new central compound/flask.}
    \label{fig:heron_scaled_concentrations}
\end{figure}

The final reaction network explored with GFN2-xTB was analyzed visually with the program \textsc{Scine}-\textsc{Heron}\cite{Bensberg2022c}, which provides an interactive view through which the network can be visualized, as illustrated in Fig.~\ref{fig:heron_scaled_concentrations}. In this network visualization, compounds and flasks are represented as green and blue dots, respectively. The lines and arrows connecting the dots represent reactions. Since the complete network cannot be shown in one figure, only the reactions of the compound or flasks at the center are shown. The selected central species can be changed by double clicking any compound/flask circle. To further restrict the reactions to a chemically meaningful selection, they can be filtered by their reaction barrier and concentration edge flux $F_I$ [see Eq.~(\ref{eq:reaction_edge_flux})] from the latest kinetic modeling simulation. Furthermore, the size of each compound/flask circle is scaled according to its maximum concentration during the micro-kinetic modeling, and the arrows representing the reactions are scaled according to $F_I$. 

The concentration-based scaling of the compound circles in Fig.~6 is chosen so that it visually represents a large range of concentrations, spanning multiple orders of magnitude. For this purpose, a logarithmic scaling function was employed to provide visually appealing sizes for the compound and flask circles: 
\begin{align}
    s(c_\mathrm{max}) = \frac{3}{2 [1 - \log_{10}(c_\mathrm{max}/(\si{mol.L^{-1}}))/2]}~,
\end{align}
where $\log_{10}$ denotes the logarithm to the basis of $10$ and $c_\mathrm{max}$ the maximum concentration of the compound observed in the latest micro-kinetic modeling simulation.
With this functions, compounds and flasks with large concentrations $c_\mathrm{max} \approx 1~\si{mol.L^{-1}}$) are represented with circles with an increased radius by a factor $s=3/2$, while for small concentrations the circle radius is decreased significantly (\emph{e.g.}, the radius is scaled by a factor of $s=0.5$ for a concentration of $c_\mathrm{max} = 10^{-4}~\si{mol.L^{-1}}$). To avoid a situation where the circle vanishes, a scaling of at most $0.1$ is applied to the circle radius.
The arrow-widths are scaled in a similar fashion according to their concentration edge flux by a factor $a$. The scaling factor is defined as
\begin{align}
    a(F_I) =\begin{cases} \frac{1}{1 - \log_{10}(F_I/(\si{mol.L^{-1}}))/2} & 1\cdot 10^{-18}\si{mol.L^{-1}} < F_I < 1.0~\si{mol.L^{-1}}\\
                            0.1 & F_I \leq 1\cdot 10^{-18}\si{mol.L^{-1}}\\
                              1.0 & \text{otherwise}
            \end{cases}~.
\end{align}

Through this interactive view, a reaction path was selected manually and refined as described in Sec.~\ref{sec:double_ended_refinement} with PBE-D3/def2-SVP/COSMO(acetonitrile) as the electronic structure model. For each selected reaction, a maximum of 10 elementary steps with the lowest transition state energy were refined. The Gibbs free energy correction for rotational, translational, and vibrational degrees of freedom for the stationary points from the refined elementary steps was then approximated with the standard rigid-rotor/particle-in-a-box/harmonic-oscillator model at a temperature of $298.15~\si{K}$ and a concentration of $1~\si{mol.L^{-1}}$. 
Furthermore, accurate electronic energies were calculated for all DFT-refined stationary points with \textsc{TightPNO}-DLPNO-CCSD(T)\cite{Riplinger2016, Riplinger2013a}, the def2-TZVP basis set~\cite{Ahlrich2005}, the conductor-like polarizable continuum model (C-PCM)\cite{Barone1998} modeling the electrostatic interaction with the solvent acetonitrile, and the quantum chemistry program \textsc{Orca} (version 4.2.1)\cite{Neese2017}. The final free energies reported in Sec.~\ref{sec:proline_catalyst_michael_addition} are the sum of the DLPNO-CCSD(T) electronic energy (including the electrostatic contribution from C-PCM) and the Gibbs free energy correction. In the following sections, we denote this combination of electronic structure models for the approximation of the Gibbs free energy as DLPNO-CCSD(T)//PBE-D3(BJ).

\section{Results}\label{sec:results}

\subsection{Micro-Kinetic Modeling-Steered Exploration}

\begin{figure}
    \centering
    \includegraphics[width=0.9\textwidth]{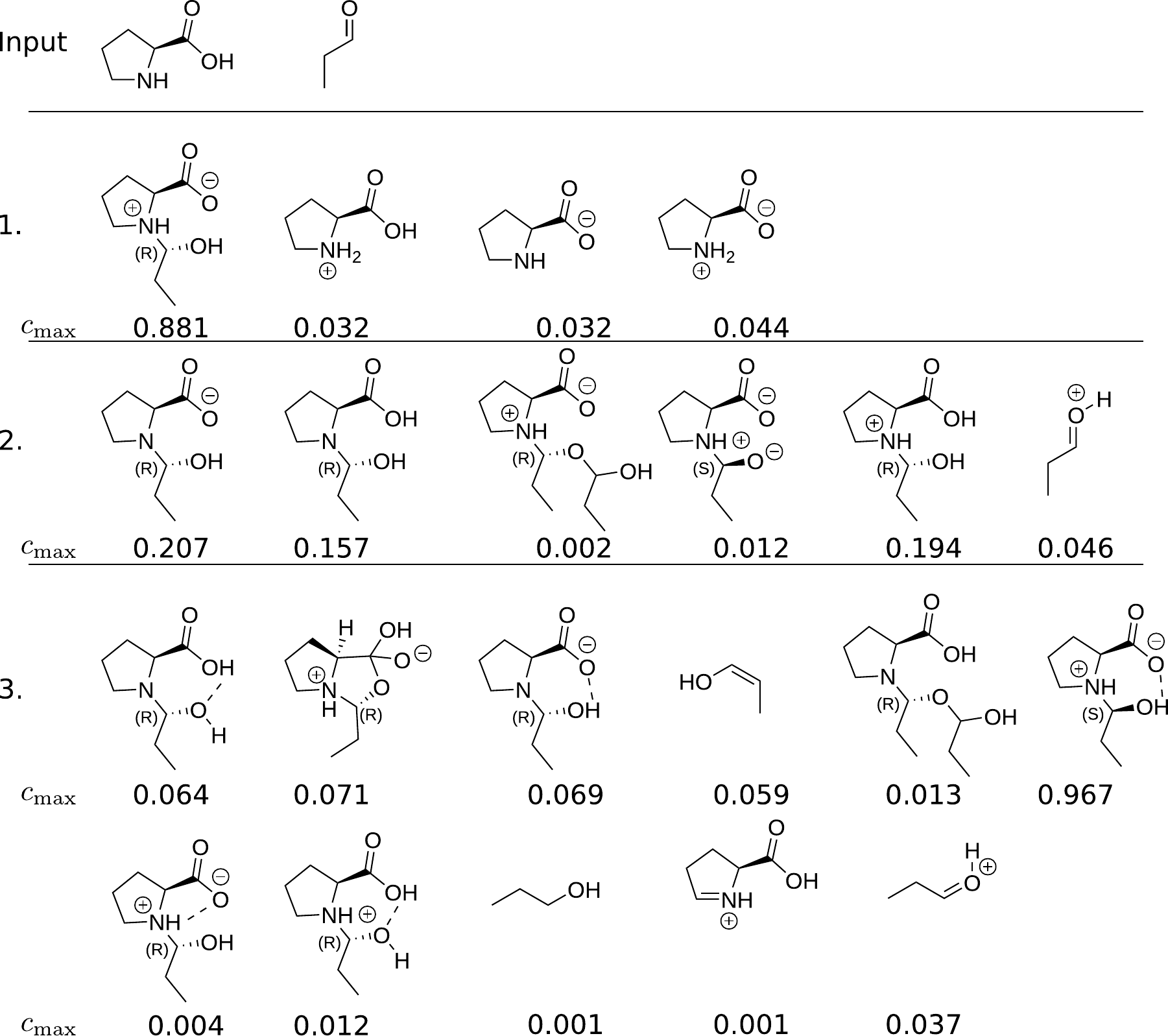}
    \caption{Compounds discovered during the first three steps of the concentration flux-driven KIEA exploration of the enamine formation. The maximum concentration obtained by kinetic modeling simulation, in which a concentration larger than $1\cdot 10^{-3}~\si{mol.L^{-1}}$ was first obtained, is provided below the Lewis structures. 
    For each step, we only show the compounds not already shown in the preceding step.
}
    \label{fig:enamine_discovered_species}
\end{figure}

To illustrate the step-wise exploration procedure of KIEA, we show the initial three sets of compounds with a maximum concentration of more than $1\cdot 10^{-3}$ discovered during the exploration of the enamine formation (step I) in Fig.~\ref{fig:enamine_discovered_species}. The exploration started with proline and propanal. After the first set of reaction trials was run and the micro-kinetic modeling was performed, the zwitterionic addition product of proline and propanal, and the three additional protonation states (zwitterionic, protonated at the nitrogen, deprotonated at the acid group) of proline were found. In the second and third steps, the different protonation states of the proline--propanal adduct and their enantiomers were formed. Most noteworthy is that a new reaction path to form the $S$ enantiomer of the adduct was obtained in the third step, leading to its large concentration.

This exploration was continued up to step 15. In this last step, no newly discovered compound fulfilled the exploration criteria given in Sec.~\ref{sec:comp_details}, and the exploration terminated. The imine (see Fig.~\ref{fig:enamine_formation}) was discovered in the fifth step. The cis- and trans-enamine 
reached a concentration of more than $1\cdot 10^{-3}$ after the 10th micro-kinetic modeling step, that is, after the 10th iteration step of reaction network exploration followed by the micro-kinetic modeling analysis of the reaction network.

The Michael addition of the enamine and nitropropene (step II), 
the water addition to the dihydrooxazine-N-oxides \textbf{dx} (step III), and the final dissociation to the product and proline (step IV) were explored in the same way. The explorations for the steps II, III, and IV converged after 7, 16, and 7 micro-kinetic modeling steps, respectively. 
All three intermediates (\textbf{zw}, \textbf{dx}, and \textbf{cb}) were discovered during the exploration and automatically probed for hydrolysis paths.

\begin{figure}
    \centering
    \includegraphics[width=0.8\textwidth]{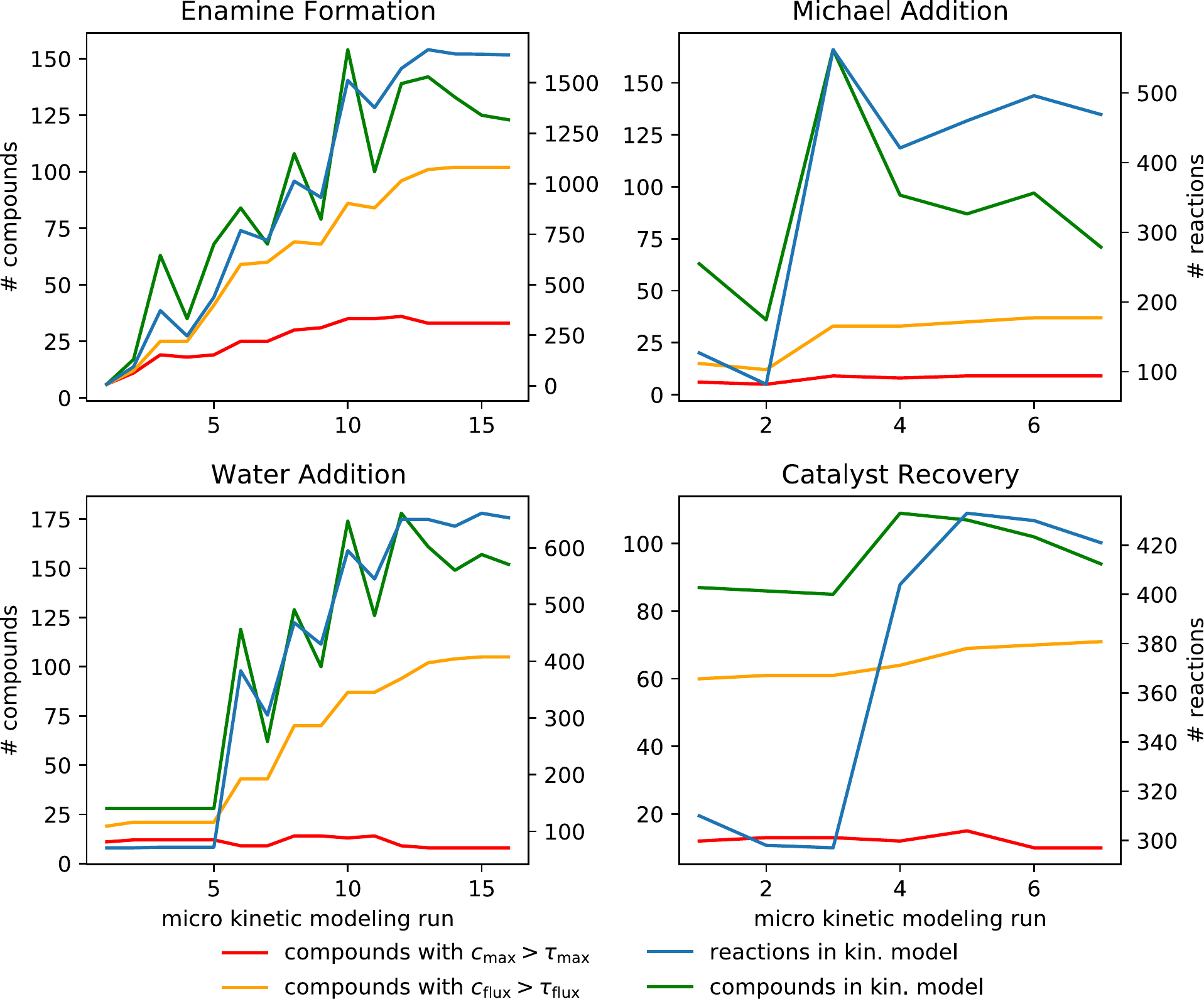}
    \caption{The number of reactions, compounds, compounds with $c^n_\mathrm{max} > \tau_\mathrm{max}$, and compounds with $c_\mathrm{flux} > \tau_\mathrm{flux}$ for
    every micro-kinetic modeling during the exploration of the four exploration steps. Compounds with $c_\mathrm{max} > \tau_\mathrm{max}$ were eligible for
    bimolecular reactions. Compounds with $c^n_\mathrm{flux} > \tau_\mathrm{flux}$ were eligible for unimolecular reactions.}
    \label{fig:kinetic_modeling_runs}
\end{figure}

The total number of reactions, compounds, and compounds with significant concentration flux ($c^n_\mathrm{flux} > \tau_\mathrm{flux}$) and significant maximum concentration ($c^n_\mathrm{max} \cdot \si{mol.L^{-1}} > \tau_\mathrm{max}$) for the micro-kinetic modeling performed during the four exploration steps are shown in Fig.~\ref{fig:kinetic_modeling_runs}. Only compounds $n$ with $c^n_\mathrm{max} \cdot \si{mol.L^{-1}} > \tau_\mathrm{max}$ can fulfill the screening condition $c_\mathrm{max}^{n} c_\mathrm{max}^{m} > \tau_\mathrm{max}$ during the exploration, and bimolecular reactions between $n$ and $m$ may be probed. The number of these compounds remained very low during the exploration. For instance, in the last micro-kinetic modeling simulation of the enamine formation, only 33  out of 123 compounds showed maximum concentrations of more than $1\cdot 10^{-2}~\si{mol.L^{-1}}$. By contrast, significantly more compounds (102) featured a large concentration flux ($c^n_\mathrm{flux} > \tau_\mathrm{flux}$) and were considered for unimolecular reactions rather than for bimolecular reactions. However, the number of reaction trials scaled linearly with the number of compounds considered for unimolecular reactions during the exploration. By contrast, the number of bimolecular reactions will scale quadratically if bimolecular reactions are not restricted, making exploring all unimolecular reactions significantly more affordable in terms of computational cost.

Because the thresholds $\tau_\mathrm{max}$ and $\tau_\mathrm{flux}$ were chosen ten times higher for the exploration steps II-IV, a significantly smaller number of compounds was considered for unimolecular and bimolecular reaction trials. Nevertheless, the total number of compounds found by the exploration, and therefore, considered in the kinetic modeling, remained comparable to the exploration of the enamine formation. During the exploration of the enamine formation, a total number of 288 compounds were considered at some point in the kinetic modeling, 266 were considered for the exploration of the Michael addition, 332 for the exploration of the water addition, and 134 for the exploration of the catalyst recovery.

During the enamine exploration, a total of 2181 reactions were considered in the micro-kinetic modeling, of which 543 were eliminated from the exploration because they had a negligible concentration flux, as described in Sec.~\ref{sec:micro-kinetic_modeling}. Therefore, the number of reactions showed strong oscillations between micro-kinetic modeling runs and did not increase monotonically. To demonstrate that our concentration-flux-based truncation scheme for the micro-kinetic modeling in KIEA introduced no significant errors, we reran the micro-kinetic modeling simulation for the 6th exploration step of the water elimination exploration step with all reactions omitted that showed a concentration flux of $1\cdot 10^{-5}$. This prescreening eliminated 81 of 383 reactions from the calculation. The maximum concentrations and final concentrations showed only insignificant differences. They changed by at most $4.6\cdot 10^{-5}~\si{mol.L^{-1}}$ and $1.62\cdot 10^{-5}~\si{mol.L^{-1}}$, respectively.

The four exploration steps uncovered a large reaction network consisting of more than $215.8~\si{k}$ elementary steps, sorted into $36.3~\si{k}$ unique reactions, $15.0~\si{k}$ compounds, $10.5~\si{k}$ flasks, and $528.0~\si{k}$ structures. Exploring the network required a total of $655.3~\si{k}$ Newton trajectory-type\cite{Unsleber2022} reaction trial calculations. While the number of exploration calculations and discovered reactions was large, only a much smaller number were considered during the micro-kinetic modeling (see Fig.~\ref{fig:kinetic_modeling_runs}). Most reactions were excluded from the micro-kinetic modeling because of their high reaction barrier. The large number of discarded reactions shows that more sophisticated algorithms identifying kinetically relevant reaction paths can significantly accelerate the exploration by reducing the number of reaction trials.

\subsection{Proline-catalyzed Michael-Addition\label{sec:proline_catalyst_michael_addition}}

\begin{figure}
    \centering
    \includegraphics[width=0.9\textwidth]{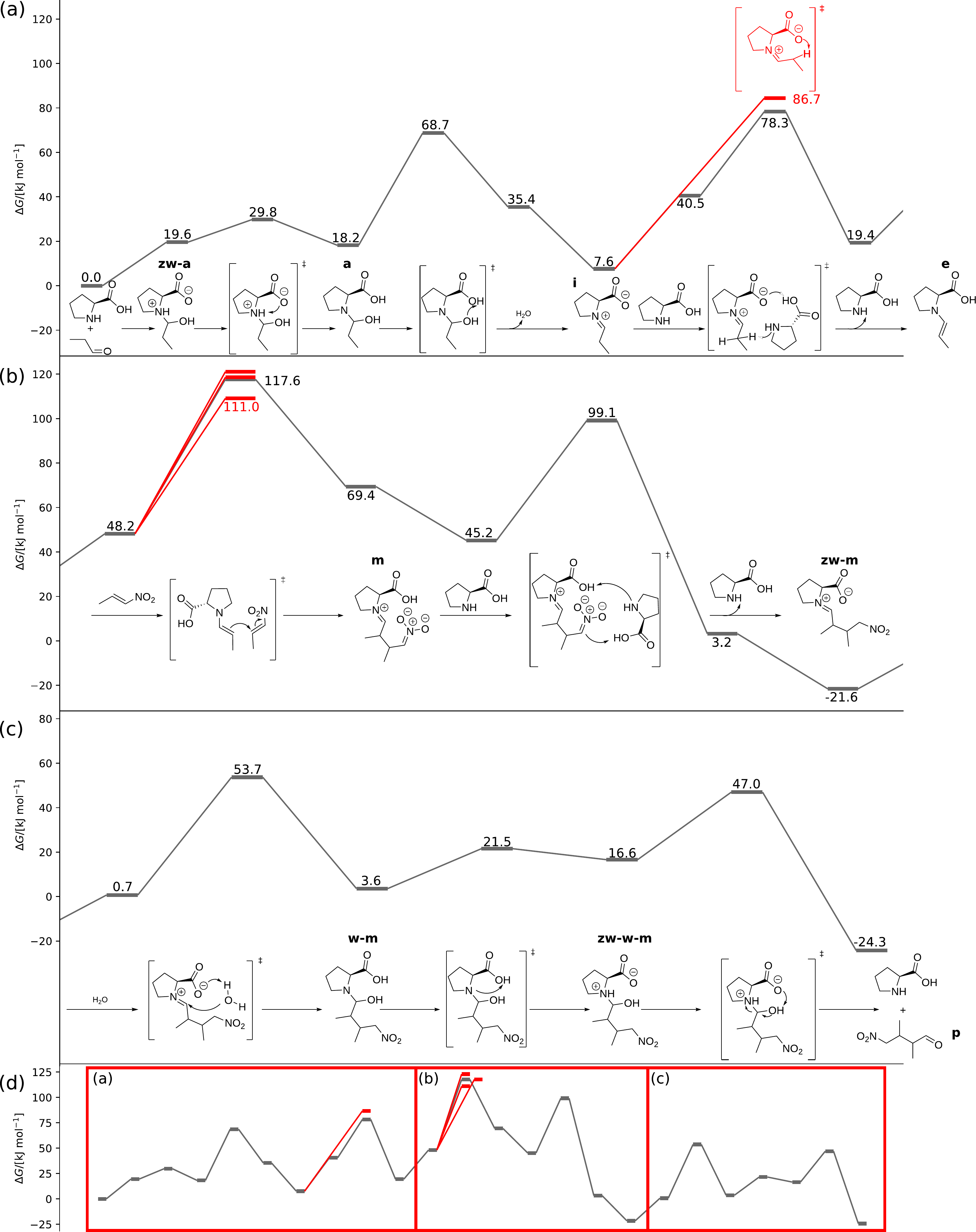}
    \caption{The potential energy diagram for the proline-catalyzed Michael addition. The complete diagram is shown in (d). The subsections for the enamine formation (a), Michael addition (b), and hydrolysis of the intermediate (c) are shown in greater detail, as indicated in (d). For \textbf{m} (and following species), only the S, R diastereomer (see Fig.~\ref{fig:michael_addition_enantiomers}) was explored. All energies are calculated with DLPNO-CCSD(T)//PBE-D3(BJ) and given in $\si{kJ.mol^{-1}}$.}
    \label{fig:full_energy_landscape}
\end{figure}

The potential energy diagram with the corresponding reaction scheme for the discovered and refined proline-catalyst reaction mechanism is shown in Fig.~\ref{fig:full_energy_landscape}. The free energies shown in Fig.~\ref{fig:full_energy_landscape} and discussed in this section were calculated DLPNO-CCSD(T)//PBE-D3(BJ). Only overall neutral compounds are depicted in the mechanism, despite the fact that charged species were explicitly considered in the automated exploration, 
as shown in Fig.~\ref{fig:enamine_discovered_species}. However, we found that the dissociation of an overall neutral compound to multiple charged compounds was strongly disfavored in terms of its free energy in acetonitrile and is therefore not considered in the reaction mechanism. We found that the nucleophilic addition of proline and propanal forming the zwitterionic intermediate \textbf{zw-a} [see Fig.~\ref{fig:full_energy_landscape}(a)] proceeds barrier-less on the PBE-D3 potential energy surface.
The reaction is favored in terms of the DLPNO-CCSD(T) electronic energies $\Delta E = -37.0~\si{kJ.mol^{-1}}$ 
but is disfavored in its DLPNO-CCSD(T)//PBE-D3(BJ) free energy change $\Delta G = 19.6~\si{kJ.mol^{-1}}$. The zwitterion \textbf{zw-a} then neutralizes its charge separation rapidly through intramolecular deprotonation of the ammonium group through the deprotonated carboxyl group, forming the intermediate \textbf{a}. We also
refined the proline-mediated protonation and deprotonation of \textbf{zw-a} to form \textbf{a}. However, the association of proline to \textbf{zw-a} to form the reacting complex was strongly disfavored ($\Delta G = 20.8~\si{kJ.mol^{-1}}$) because of its entropy reduction, making this path unlikely.

The carboxyl group of \textbf{a} protonates the OH group of \textbf{a}, leading to the dissociation of water and formation of the imine \textbf{i}. The imine \textbf{i} then forms the enamine \textbf{e} either through intramolecular deprotonation [highlighted in red in Fig.~\ref{fig:full_energy_landscape}(a)] with a DLPNO-CCSD(T)//PBE-D3(BJ) free energy barrier
of $\Delta G^\ddagger = 79.1~\si{kJ.mol^{-1}}$ compared to the \textbf{i} (free energy difference between \textbf{i} and the following transition state colored in red) or through intermolecular deprotonation through proline (association of proline: $\Delta G = 32.9~\si{kJ.mol^{-1}}$, deprotonation of the carbon atom, and protonation of the carboxyl group: $\Delta G^\ddagger = 37.9~\si{kJ.mol^{-1}}$) with an effective DLPNO-CCSD(T)//PBE-D3(BJ) barrier of
$\Delta G^\ddagger = 70.7~\si{kJ.mol^{-1}}$ (energy difference between \textbf{i} and the following transition state colored in black). Since the barriers differ by only $8.4~\si{kJ.mol^{-1}}$, we cannot rule out the intramolecular
path.
In addition, the free-energy differences are independent of the concentration of proline, which we would expect to be low during the reaction.

\begin{figure}
    \centering
    \includegraphics[width=\textwidth]{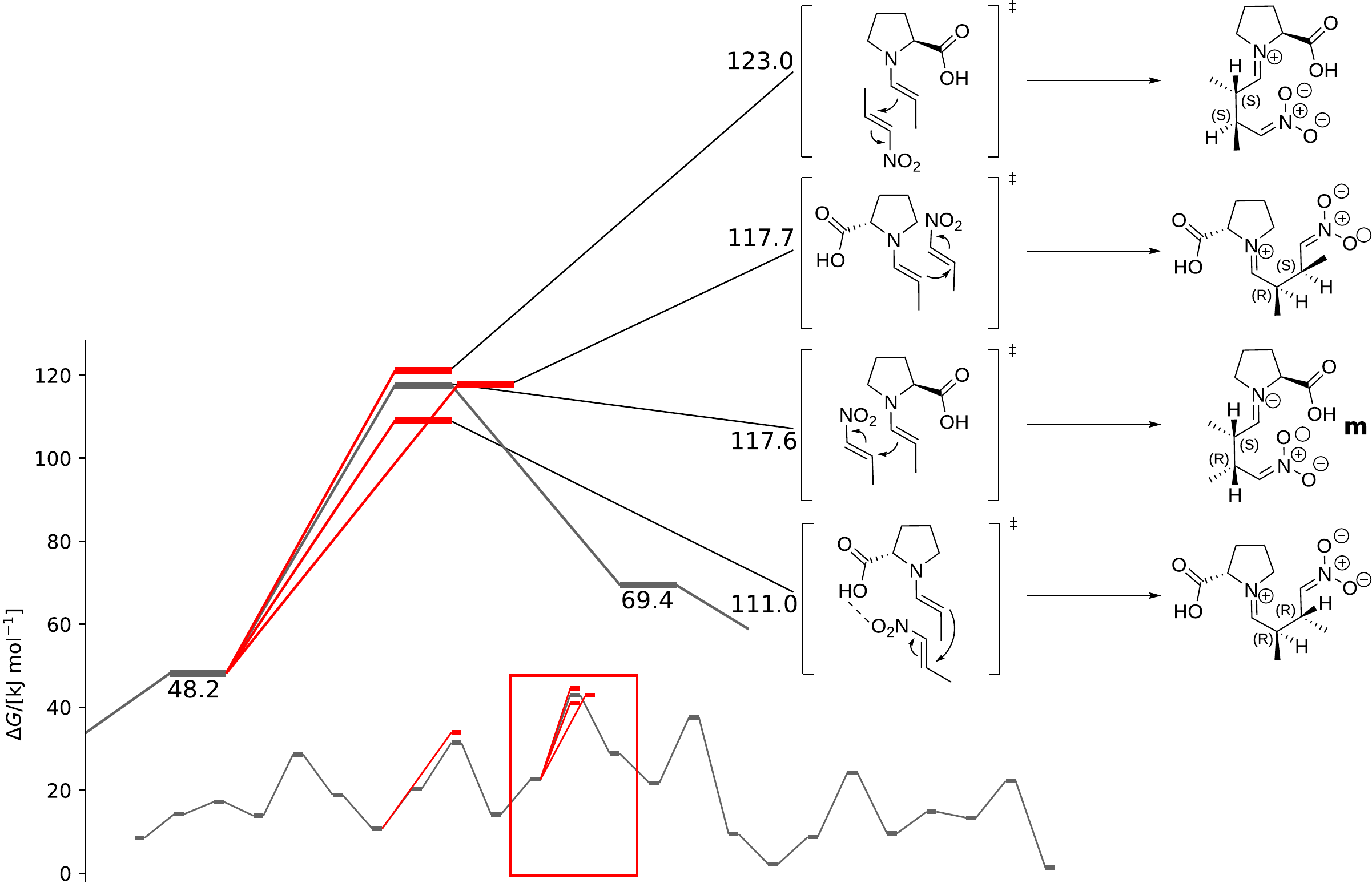}
    \caption{The transition states for the Michael addition step corresponding to different diastereomers and their DLPNO-CCSD(T)//PBE-D3(BJ) energies compared to the DLPNO-CCSD(T)//PBE-D3(BJ) energy of the separated reactants. The sketch in the bottom left indicates the position of the reaction step in the overall reaction mechanism. All energies in $\si{kJ.mol^{-1}}$.}
    \label{fig:michael_addition_enantiomers}
\end{figure}

In the next step [see Fig.~\ref{fig:full_energy_landscape}(b)], nitropropene and the enamine \textbf{e} form the reactive complex of the actual Michael addition.
Since association reactions are entropically disfavored, this step leads to a free energy increase of $\Delta G = 28.8~\si{kJ.mol^{-1}}$ to a DLPNO-CCSD(T)//PBE-D3(BJ) free energy of $48.2~\si{kJ.mol^{-1}}$ compared to the starting reactants. The Michael addition comes with a moderate DLPNO-CCSD(T)//PBE-D3(BJ) barrier of $\Delta G^\ddagger = 62.8~\si{kJ.mol^{-1}}$ compared to the reactive complex for the R,R diastereomer, $\Delta G^\ddagger = 69.4~\si{kJ.mol^{-1}}$ for the S,R diastereomer, $\Delta G^\ddagger = 69.5~\si{kJ.mol^{-1}}$ for the R,S diastereomer, and $\Delta G^\ddagger = 74.8~\si{kJ.mol^{-1}}$ for the S,S diastereomer.
A detailed comparison of the reaction barriers for the different diastereomers is shown in Fig.~\ref{fig:michael_addition_enantiomers}. After the reaction refinement, we found only for the R,R diastereomer formation a transition state in which the OH group functions as a directing group. For all other diastereomers, such transition states were not located by the automated algorithm on the DFT PES or were higher in energy. However, the short kinetic modeling runs for the Michael addition carried out during the exploration significantly favored the S,R diastereomers. Therefore, the exploration was performed only with diastereomer \textbf{m}. Note that these kinetic modeling runs were based on GFN2-xTB structures and PBE-D3 energies and not on the refined energies reported above because we explored the reaction network before we selected and refined individual reactions.
We also considered the Michael addition starting from the cis-enamine. However, all transition states located starting from the cis-enamine were higher in energy than for the trans-enamine.

Our automated reaction exploration did not find the intramolecular protonation of the CH-group of \textbf{S-m} through its carboxyl group. This failure was caused by the Newton-trajectory method implemented in the \textsc{Scine} framework not locating a suitable transition state guess. All elementary step trials targeting this reaction coordinate led to protonation of the nitro group instead. However, the proline-catalyzed protonation of the CH group and deprotonation of the carboxyl group was discovered and is shown in Fig.~\ref{fig:full_energy_landscape}(b). The carboxyl group protonates the nitrogen atom of proline while at the same time, the carboxyl group of proline protonates the CH group to which the nitro group is bonded ($\Delta G^\ddagger = 53.9~\si{kJ.mol^{-1}}$ compared to the reactive flask).
These protonation steps lead to the desired zwitterionic intermediate \textbf{zw-m}. In the following step, water coordinates to \textbf{zw-m} ($\Delta G = 22.3~\si{kJ.mol^{-1}}$), forming a reactive complex that forms the water adduct \textbf{w-m} (barrier $\Delta G^\ddagger = 53.0~\si{kJ.mol^{-1}}$ compared to the reactive flask).

The water adduct \textbf{w-m} is then brought into its zwitterionic form \textbf{zw-w-m} through intramolecular protonation ($\Delta G^\ddagger = 17.9~\si{kJ.mol^{-1}}$). We also refined the intermolecular protonation/deprotonation path with proline as a Br{\o}nsted acid catalyst. While the DLPNO-CCSD(T)//PBE-D3(BJ) reaction barrier starting from the proline-\textbf{w-m} reactive complex is lowered slightly to $\Delta G^\ddagger = 14.7~\si{kJ.mol^{-1}}$, the free energy penalty for the association of proline turned out to be too large to be competitive ($\Delta G = 54.9~\si{kJ.mol^{-1}}$).

In the last reaction step, the carbon--nitrogen bond is broken to give the final product \textbf{p}, and the catalyst proline is regenerated. This final step required only traversing a small DLPNO-CCSD(T)//PBE-D3(BJ) barrier of $\Delta G^\ddagger = 30.4~\si{kJ.mol^{-1}}$ compared to the DLPNO-CCSD(T)//PBE-D3(BJ) free energy of \textbf{zw-w-m}.

The overall reaction is found to be exergonic with a DLPNO-CCSD(T)//PBE-D3(BJ) Gibbs free energy change of $\Delta G_\mathrm{R} = -24.3~\si{kJ.m/l^{-1}}$.
The largest free energy span\cite{Uhe2010} encountered in this mechanism is $\Delta G_\mathrm{span} = 117.6~\si{kJ.mol^{-1}}$,
which is the energy difference between the starting reactants and the transition state of the Michael addition, which is reasonable for a reaction proceeding at temperatures higher than room temperature.

\section{Conclusions\label{sec:conclusion}}

We demonstrated how automated reaction network exploration can be steered by interlaced micro-kinetic modeling simulations in our KIEA approach that automatically focuses on kinetically relevant reaction paths. In addition, the concentrations calculated by the micro-kinetic modeling allowed for the visual analysis of the reaction network through the graphical user interface \textsc{Heron}, facilitating the selection of viable reaction paths for further refinement trivial.

With this highly automated approach, we explored the proline-catalyzed Michael addition of propanal and nitropropene and obtained an extensive reaction network with more than $36~\si{k}$ reactions, $25~\si{k}$ compounds and flasks. The molecular structures for this exploration were calculated with the computationally efficient GFN2-xTB approach. To obtain more reliable structures and energies, we selected a catalytic cycle of the Michael addition, re-explored the selected reactions with PBE-D3(BJ), and calculated accurate electronic energies with \textsc{TightPNO}-DLPNO-CCSD(T). The refined mechanism is non-trivial, containing 14 reaction steps, connecting a total of 24 stationary points with no unreasonably high reaction barriers, suggesting that the reaction should proceed at temperatures slightly above room temperature.

The largest molecular complex in the refined catalytic cycle contained 52 atoms. The size of the molecular systems studied in this work clearly demonstrates that automated reaction network exploration is not limited to small molecular systems but leverages in-depth investigations into reaction mechanisms.

The extensive GFN2-xTB-based network allowed us to investigate competing reaction paths quickly. For instance, we found that the deprotonation of imine \textbf{i} to form enamine \textbf{e} is accelerated by a second proline molecule, performing the deprotonation through an intermolecular instead of an intramolecular mechanism. Furthermore, the network directly provided all four possible diastereomers formed in the Michael addition step between the enamine \textbf{e} and nitropropene. 

The automated exploration based on GFN2-xTB considered the S,R diastereomer as the kinetically most relevant, leading to its further exploration. The refinement revealed that the R,R diastereomer is kinetically favored by $6.6~\si{kJ.mol^{-1}}$. This could be addressed in the future by interlocking the automated
refinement of the reactions through a more accurate electronic structure method for structures and energies by refining the reaction steps already during the exploration. For instance, the kinetic modeling could initially identify relevant reactions for the refinement. New compounds could then be selected for further exploration based on the refined reactions. Such an interlocked approach should be able to directly map between compounds found with the different electronic structure methods and resolve the challenge that not every compound may exist as a stationary point for the different electronic structure models, as discussed in Sec.~\ref{sec:double_ended_refinement}.
Furthermore, the exploration could be massively accelerated through more sophisticated approaches, \emph{e.g.}, data-driven approaches that aim at extracting and generalizing promising reaction coordinates from short but unconstrained automated reaction network explorations\cite{Unsleber2023},
selecting reaction coordinates for reaction-trial probing that lead to kinetically relevant reactions, reducing the number of reaction trials by more than one order of magnitude.

\section*{Data Availability}
 The reaction network exploration data is available through the PROLINE Exploration Data Set\cite{Bensberg2023c}.

\providecommand{\latin}[1]{#1}
\makeatletter
\providecommand{\doi}
  {\begingroup\let\do\@makeother\dospecials
  \catcode`\{=1 \catcode`\}=2 \doi@aux}
\providecommand{\doi@aux}[1]{\endgroup\texttt{#1}}
\makeatother
\providecommand*\mcitethebibliography{\thebibliography}
\csname @ifundefined\endcsname{endmcitethebibliography}
  {\let\endmcitethebibliography\endthebibliography}{}


\begin{mcitethebibliography}{81}
\providecommand*\natexlab[1]{#1}
\providecommand*\mciteSetBstSublistMode[1]{}
\providecommand*\mciteSetBstMaxWidthForm[2]{}
\providecommand*\mciteBstWouldAddEndPuncttrue
  {\def\EndOfBibitem{\unskip.}}
\providecommand*\mciteBstWouldAddEndPunctfalse
  {\let\EndOfBibitem\relax}
\providecommand*\mciteSetBstMidEndSepPunct[3]{}
\providecommand*\mciteSetBstSublistLabelBeginEnd[3]{}
\providecommand*\EndOfBibitem{}
\mciteSetBstSublistMode{f}
\mciteSetBstMaxWidthForm{subitem}{(\alph{mcitesubitemcount})}
\mciteSetBstSublistLabelBeginEnd
  {\mcitemaxwidthsubitemform\space}
  {\relax}
  {\relax}

\bibitem[Eyring(1935)]{Eyring1935}
Eyring,~H. The Activated Complex in Chemical Reactions. \emph{J. Chem. Phys.}
  \textbf{1935}, \emph{3}, 107--115\relax
\mciteBstWouldAddEndPuncttrue
\mciteSetBstMidEndSepPunct{\mcitedefaultmidpunct}
{\mcitedefaultendpunct}{\mcitedefaultseppunct}\relax
\EndOfBibitem
\bibitem[Truhlar \latin{et~al.}(1996)Truhlar, Garrett, and
  Klippenstein]{Truhlar1996}
Truhlar,~D.~G.; Garrett,~B.~C.; Klippenstein,~S.~J. Current Status of
  Transition-State Theory. \emph{J. Phys. Chem.} \textbf{1996}, \emph{100},
  12771--12800\relax
\mciteBstWouldAddEndPuncttrue
\mciteSetBstMidEndSepPunct{\mcitedefaultmidpunct}
{\mcitedefaultendpunct}{\mcitedefaultseppunct}\relax
\EndOfBibitem
\bibitem[Sameera \latin{et~al.}(2016)Sameera, Maeda, and Morokuma]{Sameera2016}
Sameera,~W. M.~C.; Maeda,~S.; Morokuma,~K. Computational Catalysis Using the
  Artificial Force Induced Reaction Method. \emph{Accounts Chem. Res.}
  \textbf{2016}, \emph{49}, 763--773\relax
\mciteBstWouldAddEndPuncttrue
\mciteSetBstMidEndSepPunct{\mcitedefaultmidpunct}
{\mcitedefaultendpunct}{\mcitedefaultseppunct}\relax
\EndOfBibitem
\bibitem[Dewyer \latin{et~al.}(2017)Dewyer, Argüelles, and
  Zimmerman]{Dewyer2017}
Dewyer,~A.~L.; Argüelles,~A.~J.; Zimmerman,~P.~M. Methods for exploring
  reaction space in molecular systems. \emph{WIREs Comput Mol Sci}
  \textbf{2017}, \emph{8}, e1354\relax
\mciteBstWouldAddEndPuncttrue
\mciteSetBstMidEndSepPunct{\mcitedefaultmidpunct}
{\mcitedefaultendpunct}{\mcitedefaultseppunct}\relax
\EndOfBibitem
\bibitem[V{\'a}zquez \latin{et~al.}(2018)V{\'a}zquez, Otero, and
  Mart{\'i}nez-N{\'u}{\~n}ez]{Vazquez2018}
V{\'a}zquez,~S.~A.; Otero,~X.~L.; Mart{\'i}nez-N{\'u}{\~n}ez,~E. {A
  Trajectory-Based Method to Explore Reaction Mechanisms}. \emph{Molecules}
  \textbf{2018}, \emph{23}, 3156\relax
\mciteBstWouldAddEndPuncttrue
\mciteSetBstMidEndSepPunct{\mcitedefaultmidpunct}
{\mcitedefaultendpunct}{\mcitedefaultseppunct}\relax
\EndOfBibitem
\bibitem[Simm \latin{et~al.}(2019)Simm, Vaucher, and Reiher]{Simm2018}
Simm,~G.~N.; Vaucher,~A.~C.; Reiher,~M. Exploration of Reaction Pathways and
  Chemical Transformation Networks. \emph{J. Phys. Chem. A} \textbf{2019},
  \emph{123}, 385--399\relax
\mciteBstWouldAddEndPuncttrue
\mciteSetBstMidEndSepPunct{\mcitedefaultmidpunct}
{\mcitedefaultendpunct}{\mcitedefaultseppunct}\relax
\EndOfBibitem
\bibitem[Green(2019)]{Green2019}
Green,~W.~H. \emph{Computer Aided Chemical Engineering}; Elsevier, 2019; pp
  259--294\relax
\mciteBstWouldAddEndPuncttrue
\mciteSetBstMidEndSepPunct{\mcitedefaultmidpunct}
{\mcitedefaultendpunct}{\mcitedefaultseppunct}\relax
\EndOfBibitem
\bibitem[Unsleber and Reiher(2020)Unsleber, and Reiher]{Unsleber2020}
Unsleber,~J.~P.; Reiher,~M. The Exploration of Chemical Reaction Networks.
  \emph{Annu. Rev. Phys. Chem.} \textbf{2020}, \emph{71}, 121--142\relax
\mciteBstWouldAddEndPuncttrue
\mciteSetBstMidEndSepPunct{\mcitedefaultmidpunct}
{\mcitedefaultendpunct}{\mcitedefaultseppunct}\relax
\EndOfBibitem
\bibitem[Maeda and Harabuchi(2021)Maeda, and Harabuchi]{Maeda2021}
Maeda,~S.; Harabuchi,~Y. {Exploring Paths of Chemical Transformations in
  Molecular and Periodic Systems: An Approach Utilizing Force}. \emph{WIREs
  Comput. Mol. Sci.} \textbf{2021}, e1538\relax
\mciteBstWouldAddEndPuncttrue
\mciteSetBstMidEndSepPunct{\mcitedefaultmidpunct}
{\mcitedefaultendpunct}{\mcitedefaultseppunct}\relax
\EndOfBibitem
\bibitem[Baiardi \latin{et~al.}(2021)Baiardi, Grimmel, Steiner, Türtscher,
  Unsleber, Weymuth, and Reiher]{Baiardi2021}
Baiardi,~A.; Grimmel,~S.~A.; Steiner,~M.; Türtscher,~P.~L.; Unsleber,~J.~P.;
  Weymuth,~T.; Reiher,~M. Expansive Quantum Mechanical Exploration of Chemical
  Reaction Paths. \emph{Accounts Chem. Res.} \textbf{2021}, \emph{55},
  35--43\relax
\mciteBstWouldAddEndPuncttrue
\mciteSetBstMidEndSepPunct{\mcitedefaultmidpunct}
{\mcitedefaultendpunct}{\mcitedefaultseppunct}\relax
\EndOfBibitem
\bibitem[Steiner and Reiher(2022)Steiner, and Reiher]{Steiner2022}
Steiner,~M.; Reiher,~M. Autonomous Reaction Network Exploration in Homogeneous
  and Heterogeneous Catalysis. \emph{Top. Catal.} \textbf{2022}, \emph{65},
  6--39\relax
\mciteBstWouldAddEndPuncttrue
\mciteSetBstMidEndSepPunct{\mcitedefaultmidpunct}
{\mcitedefaultendpunct}{\mcitedefaultseppunct}\relax
\EndOfBibitem
\bibitem[Susnow \latin{et~al.}(1997)Susnow, Dean, Green, Peczak, and
  Broadbelt]{Susnow1997}
Susnow,~R.~G.; Dean,~A.~M.; Green,~W.~H.; Peczak,~P.; Broadbelt,~L.~J.
  Rate-Based Construction of Kinetic Models for Complex Systems. \emph{J. Phys.
  Chem. A} \textbf{1997}, \emph{101}, 3731--3740\relax
\mciteBstWouldAddEndPuncttrue
\mciteSetBstMidEndSepPunct{\mcitedefaultmidpunct}
{\mcitedefaultendpunct}{\mcitedefaultseppunct}\relax
\EndOfBibitem
\bibitem[Broadbelt \latin{et~al.}(1994)Broadbelt, Stark, and
  Klein]{broadbelt1994computer}
Broadbelt,~L.~J.; Stark,~S.~M.; Klein,~M.~T. Computer generated pyrolysis
  modeling: on-the-fly generation of species, reactions, and rates. \emph{Ind.
  Eng. Chem. Res.} \textbf{1994}, \emph{33}, 790--799\relax
\mciteBstWouldAddEndPuncttrue
\mciteSetBstMidEndSepPunct{\mcitedefaultmidpunct}
{\mcitedefaultendpunct}{\mcitedefaultseppunct}\relax
\EndOfBibitem
\bibitem[Broadbelt \latin{et~al.}(1994)Broadbelt, Stark, and
  Klein]{Broadbelt1994}
Broadbelt,~L.~J.; Stark,~S.~M.; Klein,~M.~T. Computer generated reaction
  networks: on-the-fly calculation of species properties using computational
  quantum chemistry. \emph{Chem. Eng. Sci.} \textbf{1994}, \emph{49},
  4991--5010\relax
\mciteBstWouldAddEndPuncttrue
\mciteSetBstMidEndSepPunct{\mcitedefaultmidpunct}
{\mcitedefaultendpunct}{\mcitedefaultseppunct}\relax
\EndOfBibitem
\bibitem[Witt \latin{et~al.}(2000)Witt, Dooling, and Broadbelt]{Witt2000}
Witt,~M. J.~D.; Dooling,~D.~J.; Broadbelt,~L.~J. Computer Generation of
  Reaction Mechanisms Using Quantitative Rate Information:{\hspace{0.167em}}
  Application to Long-Chain Hydrocarbon Pyrolysis. \emph{Ind. Eng. Chem. Res.}
  \textbf{2000}, \emph{39}, 2228--2237\relax
\mciteBstWouldAddEndPuncttrue
\mciteSetBstMidEndSepPunct{\mcitedefaultmidpunct}
{\mcitedefaultendpunct}{\mcitedefaultseppunct}\relax
\EndOfBibitem
\bibitem[Koninckx \latin{et~al.}(2022)Koninckx, Colin, Broadbelt, and
  Vernuccio]{Koninckx2022}
Koninckx,~E.; Colin,~J.~G.; Broadbelt,~L.~J.; Vernuccio,~S. Catalytic
  Conversion of Alkenes on Acidic Zeolites: Automated Generation of Reaction
  Mechanisms and Lumping Technique. \emph{{ACS} Eng. Au} \textbf{2022}, \relax
\mciteBstWouldAddEndPunctfalse
\mciteSetBstMidEndSepPunct{\mcitedefaultmidpunct}
{}{\mcitedefaultseppunct}\relax
\EndOfBibitem
\bibitem[Warth \latin{et~al.}(2000)Warth, Battin-Leclerc, Fournet, Glaude,
  C{\^{o}}me, and Scacchi]{Warth2000}
Warth,~V.; Battin-Leclerc,~F.; Fournet,~R.; Glaude,~P.; C{\^{o}}me,~G.;
  Scacchi,~G. Computer based generation of reaction mechanisms for gas-phase
  oxidation. \emph{Comput. Chem.} \textbf{2000}, \emph{24}, 541--560\relax
\mciteBstWouldAddEndPuncttrue
\mciteSetBstMidEndSepPunct{\mcitedefaultmidpunct}
{\mcitedefaultendpunct}{\mcitedefaultseppunct}\relax
\EndOfBibitem
\bibitem[Gao \latin{et~al.}(2016)Gao, Allen, Green, and West]{Gao2016}
Gao,~C.~W.; Allen,~J.~W.; Green,~W.~H.; West,~R.~H. Reaction Mechanism
  Generator: Automatic construction of chemical kinetic mechanisms.
  \emph{Comput. Phys. Commun.} \textbf{2016}, \emph{203}, 212--225\relax
\mciteBstWouldAddEndPuncttrue
\mciteSetBstMidEndSepPunct{\mcitedefaultmidpunct}
{\mcitedefaultendpunct}{\mcitedefaultseppunct}\relax
\EndOfBibitem
\bibitem[Goldsmith and West(2017)Goldsmith, and West]{Goldsmith2017}
Goldsmith,~C.~F.; West,~R.~H. Automatic Generation of Microkinetic Mechanisms
  for Heterogeneous Catalysis. \emph{J. Phys. Chem. C} \textbf{2017},
  \emph{121}, 9970--9981\relax
\mciteBstWouldAddEndPuncttrue
\mciteSetBstMidEndSepPunct{\mcitedefaultmidpunct}
{\mcitedefaultendpunct}{\mcitedefaultseppunct}\relax
\EndOfBibitem
\bibitem[Liu \latin{et~al.}(2021)Liu, Dana, Johnson, Goldman, Jocher, Payne,
  Grambow, Han, Yee, Mazeau, Blondal, West, Goldsmith, and Green]{Liu2021}
Liu,~M.; Dana,~A.~G.; Johnson,~M.~S.; Goldman,~M.~J.; Jocher,~A.; Payne,~A.~M.;
  Grambow,~C.~A.; Han,~K.; Yee,~N.~W.; Mazeau,~E.~J.; Blondal,~K.; West,~R.~H.;
  Goldsmith,~C.~F.; Green,~W.~H. Reaction Mechanism Generator v3.0: Advances in
  Automatic Mechanism Generation. \emph{J. Chem. Inf. Model.} \textbf{2021},
  \emph{61}, 2686--2696\relax
\mciteBstWouldAddEndPuncttrue
\mciteSetBstMidEndSepPunct{\mcitedefaultmidpunct}
{\mcitedefaultendpunct}{\mcitedefaultseppunct}\relax
\EndOfBibitem
\bibitem[Johnson \latin{et~al.}(2022)Johnson, Dong, Dana, Chung, Farina,
  Gillis, Liu, Yee, Blondal, Mazeau, Grambow, Payne, Spiekermann, Pang,
  Goldsmith, West, and Green]{Johnson2022}
Johnson,~M.~S.; Dong,~X.; Dana,~A.~G.; Chung,~Y.; Farina,~D.; Gillis,~R.~J.;
  Liu,~M.; Yee,~N.~W.; Blondal,~K.; Mazeau,~E.; Grambow,~C.~A.; Payne,~A.~M.;
  Spiekermann,~K.~A.; Pang,~H.-W.; Goldsmith,~C.~F.; West,~R.~H.; Green,~W.~H.
  {RMG} Database for Chemical Property Prediction. \emph{J. Chem. Inf. Model.}
  \textbf{2022}, \emph{62}, 4906--4915\relax
\mciteBstWouldAddEndPuncttrue
\mciteSetBstMidEndSepPunct{\mcitedefaultmidpunct}
{\mcitedefaultendpunct}{\mcitedefaultseppunct}\relax
\EndOfBibitem
\bibitem[Xie \latin{et~al.}(2021)Xie, Spotte-Smith, Wen, Patel, Blau, and
  Persson]{Xie2021}
Xie,~X.; Spotte-Smith,~E. W.~C.; Wen,~M.; Patel,~H.~D.; Blau,~S.~M.;
  Persson,~K.~A. Data-Driven Prediction of Formation Mechanisms of Lithium
  Ethylene Monocarbonate with an Automated Reaction Network. \emph{J. Am. Chem.
  Soc.} \textbf{2021}, \emph{143}, 13245--13258\relax
\mciteBstWouldAddEndPuncttrue
\mciteSetBstMidEndSepPunct{\mcitedefaultmidpunct}
{\mcitedefaultendpunct}{\mcitedefaultseppunct}\relax
\EndOfBibitem
\bibitem[Blau \latin{et~al.}(2021)Blau, Patel, Spotte-Smith, Xie, Dwaraknath,
  and Persson]{Blau2021}
Blau,~S.~M.; Patel,~H.~D.; Spotte-Smith,~E. W.~C.; Xie,~X.; Dwaraknath,~S.;
  Persson,~K.~A. A chemically consistent graph architecture for massive
  reaction networks applied to solid-electrolyte interphase formation.
  \emph{Chem. Sci.} \textbf{2021}, \emph{12}, 4931--4939\relax
\mciteBstWouldAddEndPuncttrue
\mciteSetBstMidEndSepPunct{\mcitedefaultmidpunct}
{\mcitedefaultendpunct}{\mcitedefaultseppunct}\relax
\EndOfBibitem
\bibitem[Türtscher and Reiher(2022)Türtscher, and Reiher]{Tuertscher2022}
Türtscher,~P.~L.; Reiher,~M. Pathfinder---Navigating and Analyzing Chemical
  Reaction Networks with an Efficient Graph-based Approach. \emph{J. Chem. Inf.
  Model.} \textbf{2022}, doi.org/10.1021/acs.jcim.2c01136\relax
\mciteBstWouldAddEndPuncttrue
\mciteSetBstMidEndSepPunct{\mcitedefaultmidpunct}
{\mcitedefaultendpunct}{\mcitedefaultseppunct}\relax
\EndOfBibitem
\bibitem[Proppe \latin{et~al.}(2016)Proppe, Husch, Simm, and
  Reiher]{Proppe2016}
Proppe,~J.; Husch,~T.; Simm,~G.~N.; Reiher,~M. Uncertainty quantification for
  quantum chemical models of complex reaction networks. \emph{Faraday Discuss.}
  \textbf{2016}, \emph{195}, 497--520\relax
\mciteBstWouldAddEndPuncttrue
\mciteSetBstMidEndSepPunct{\mcitedefaultmidpunct}
{\mcitedefaultendpunct}{\mcitedefaultseppunct}\relax
\EndOfBibitem
\bibitem[Proppe and Reiher(2018)Proppe, and Reiher]{Proppe2018}
Proppe,~J.; Reiher,~M. Mechanism Deduction from Noisy Chemical Reaction
  Networks. \emph{J. Chem. Theory Comput.} \textbf{2018}, \emph{15},
  357--370\relax
\mciteBstWouldAddEndPuncttrue
\mciteSetBstMidEndSepPunct{\mcitedefaultmidpunct}
{\mcitedefaultendpunct}{\mcitedefaultseppunct}\relax
\EndOfBibitem
\bibitem[Sumiya and Maeda(2020)Sumiya, and Maeda]{Sumiya2020}
Sumiya,~Y.; Maeda,~S. Rate Constant Matrix Contraction Method for Systematic
  Analysis of Reaction Path Networks. \emph{Chem. Lett.} \textbf{2020},
  \emph{49}, 553--564\relax
\mciteBstWouldAddEndPuncttrue
\mciteSetBstMidEndSepPunct{\mcitedefaultmidpunct}
{\mcitedefaultendpunct}{\mcitedefaultseppunct}\relax
\EndOfBibitem
\bibitem[Matheu \latin{et~al.}(2001)Matheu, Lada, Green, Dean, and
  Grenda]{Matheu2001}
Matheu,~D.~M.; Lada,~T.~A.; Green,~W.~H.; Dean,~A.~M.; Grenda,~J.~M.
  {Rate-based screening of pressure-dependent reaction networks}.
  \emph{{Comput. Phys. Commun.}} \textbf{2001}, \emph{138}, 237--249\relax
\mciteBstWouldAddEndPuncttrue
\mciteSetBstMidEndSepPunct{\mcitedefaultmidpunct}
{\mcitedefaultendpunct}{\mcitedefaultseppunct}\relax
\EndOfBibitem
\bibitem[Geem \latin{et~al.}(2006)Geem, Reyniers, Marin, Song, Green, and
  Matheu]{Geem2006}
Geem,~K. M.~V.; Reyniers,~M.-F.; Marin,~G.~B.; Song,~J.; Green,~W.~H.;
  Matheu,~D.~M. Automatic reaction network generation using {RMG} for steam
  cracking of n-hexane. \emph{AIChE J.} \textbf{2006}, \emph{52},
  718--730\relax
\mciteBstWouldAddEndPuncttrue
\mciteSetBstMidEndSepPunct{\mcitedefaultmidpunct}
{\mcitedefaultendpunct}{\mcitedefaultseppunct}\relax
\EndOfBibitem
\bibitem[Blurock \latin{et~al.}(2012)Blurock, Battin-Leclerc, Faravelli, and
  Green]{Blurock2012}
Blurock,~E.; Battin-Leclerc,~F.; Faravelli,~T.; Green,~W.~H. In \emph{{Cleaner
  Combustion: Developing Detailed Chemical Kinetic Models}};
  F.~Battin-Leclerc,~E.~B.,~J.~Simmie, Ed.; Springer-Verlag, 2012; Chapter
  {Automatic generation of detailed mechanisms}, pp 59--92\relax
\mciteBstWouldAddEndPuncttrue
\mciteSetBstMidEndSepPunct{\mcitedefaultmidpunct}
{\mcitedefaultendpunct}{\mcitedefaultseppunct}\relax
\EndOfBibitem
\bibitem[List \latin{et~al.}(2000)List, Lerner, and Barbas]{List2000}
List,~B.; Lerner,~R.~A.; Barbas,~C.~F. Proline-Catalyzed Direct Asymmetric
  Aldol Reactions. \emph{J. Am. Chem. Soc.} \textbf{2000}, \emph{122},
  2395--2396\relax
\mciteBstWouldAddEndPuncttrue
\mciteSetBstMidEndSepPunct{\mcitedefaultmidpunct}
{\mcitedefaultendpunct}{\mcitedefaultseppunct}\relax
\EndOfBibitem
\bibitem[Mukherjee \latin{et~al.}(2007)Mukherjee, Yang, Hoffmann, and
  List]{Mukherjee2007}
Mukherjee,~S.; Yang,~J.~W.; Hoffmann,~S.; List,~B. Asymmetric Enamine
  Catalysis. \emph{Chem. Rev.} \textbf{2007}, \emph{107}, 5471--5569\relax
\mciteBstWouldAddEndPuncttrue
\mciteSetBstMidEndSepPunct{\mcitedefaultmidpunct}
{\mcitedefaultendpunct}{\mcitedefaultseppunct}\relax
\EndOfBibitem
\bibitem[Seayad and List(2005)Seayad, and List]{Seayad2005}
Seayad,~J.; List,~B. Asymmetric organocatalysis. \emph{Org. Biomol. Chem.}
  \textbf{2005}, \emph{3}, 719\relax
\mciteBstWouldAddEndPuncttrue
\mciteSetBstMidEndSepPunct{\mcitedefaultmidpunct}
{\mcitedefaultendpunct}{\mcitedefaultseppunct}\relax
\EndOfBibitem
\bibitem[Melchiorre \latin{et~al.}(2008)Melchiorre, Marigo, Carlone, and
  Bartoli]{Melchiorre2008}
Melchiorre,~P.; Marigo,~M.; Carlone,~A.; Bartoli,~G. Asymmetric
  Aminocatalysis-Gold Rush in Organic Chemistry. \emph{Angew. Chem. Int. Ed.}
  \textbf{2008}, \emph{47}, 6138--6171\relax
\mciteBstWouldAddEndPuncttrue
\mciteSetBstMidEndSepPunct{\mcitedefaultmidpunct}
{\mcitedefaultendpunct}{\mcitedefaultseppunct}\relax
\EndOfBibitem
\bibitem[List \latin{et~al.}(2001)List, Pojarliev, and Martin]{List2001}
List,~B.; Pojarliev,~P.; Martin,~H.~J. Efficient Proline-Catalyzed Michael
  Additions of Unmodified Ketones to Nitro Olefins. \emph{Org. Lett.}
  \textbf{2001}, \emph{3}, 2423--2425\relax
\mciteBstWouldAddEndPuncttrue
\mciteSetBstMidEndSepPunct{\mcitedefaultmidpunct}
{\mcitedefaultendpunct}{\mcitedefaultseppunct}\relax
\EndOfBibitem
\bibitem[Maillard \latin{et~al.}(2019)Maillard, Park, and Kang]{Maillard2019}
Maillard,~L.~T.; Park,~H.~S.; Kang,~Y.~K. Organocatalytic Asymmetric Addition
  of Aldehyde to Nitroolefin by H-D-Pro-Pro-Glu-{NH}$_2$: A Mechanistic Study.
  \emph{ACS Omega} \textbf{2019}, \emph{4}, 8862--8873\relax
\mciteBstWouldAddEndPuncttrue
\mciteSetBstMidEndSepPunct{\mcitedefaultmidpunct}
{\mcitedefaultendpunct}{\mcitedefaultseppunct}\relax
\EndOfBibitem
\bibitem[Castro-Alvarez \latin{et~al.}(2019)Castro-Alvarez, Carneros, Calafat,
  Costa, Marco, and Vilarrasa]{CastroAlvarez2019}
Castro-Alvarez,~A.; Carneros,~H.; Calafat,~J.; Costa,~A.~M.; Marco,~C.;
  Vilarrasa,~J. {NMR} and Computational Studies on the Reactions of Enamines
  with Nitroalkenes That May Pass through Cyclobutanes. \emph{ACS Omega}
  \textbf{2019}, \emph{4}, 18167--18194\relax
\mciteBstWouldAddEndPuncttrue
\mciteSetBstMidEndSepPunct{\mcitedefaultmidpunct}
{\mcitedefaultendpunct}{\mcitedefaultseppunct}\relax
\EndOfBibitem
\bibitem[Sahoo \latin{et~al.}(2012)Sahoo, Rahaman, Madar{\'{a}}sz, P{\'{a}}pai,
  Melarto, Valkonen, and Pihko]{Sahoo2012}
Sahoo,~G.; Rahaman,~H.; Madar{\'{a}}sz,~{\'{A}}.; P{\'{a}}pai,~I.; Melarto,~M.;
  Valkonen,~A.; Pihko,~P.~M. Dihydrooxazine Oxides as Key Intermediates in
  Organocatalytic Michael Additions of Aldehydes to Nitroalkenes. \emph{Angew.
  Chem. Int. Ed.} \textbf{2012}, \emph{51}, 13144--13148\relax
\mciteBstWouldAddEndPuncttrue
\mciteSetBstMidEndSepPunct{\mcitedefaultmidpunct}
{\mcitedefaultendpunct}{\mcitedefaultseppunct}\relax
\EndOfBibitem
\bibitem[Seebach \latin{et~al.}(2013)Seebach, Sun, Ebert, Schweizer,
  Purkayastha, Beck, Duschmal{\'{e}}, Wennemers, Mukaiyama, Benohoud, Hayashi,
  and Reiher]{Seebach2013}
Seebach,~D.; Sun,~X.; Ebert,~M.-O.; Schweizer,~W.~B.; Purkayastha,~N.;
  Beck,~A.~K.; Duschmal{\'{e}},~J.; Wennemers,~H.; Mukaiyama,~T.; Benohoud,~M.;
  Hayashi,~Y.; Reiher,~M. Stoichiometric Reactions of Enamines Derived from
  Diphenylprolinol Silyl Ethers with Nitro Olefins and Lessons for the
  Corresponding Organocatalytic Conversions - a Survey. \emph{Helv. Chim. Acta}
  \textbf{2013}, \emph{96}, 799--852\relax
\mciteBstWouldAddEndPuncttrue
\mciteSetBstMidEndSepPunct{\mcitedefaultmidpunct}
{\mcitedefaultendpunct}{\mcitedefaultseppunct}\relax
\EndOfBibitem
\bibitem[Vilarrasa \latin{et~al.}(2017)Vilarrasa, Castro-Alvarez, Carneros, and
  Costa]{Vilarrasa2017}
Vilarrasa,~J.; Castro-Alvarez,~A.; Carneros,~H.; Costa,~A.~M. Computer-Aided
  Insight into the Relative Stability of Enamines. \emph{Synthesis}
  \textbf{2017}, \emph{49}, 5285--5306\relax
\mciteBstWouldAddEndPuncttrue
\mciteSetBstMidEndSepPunct{\mcitedefaultmidpunct}
{\mcitedefaultendpunct}{\mcitedefaultseppunct}\relax
\EndOfBibitem
\bibitem[Patora-Komisarska \latin{et~al.}(2011)Patora-Komisarska, Benohoud,
  Ishikawa, Seebach, and Hayashi]{PatoraKomisarska2011}
Patora-Komisarska,~K.; Benohoud,~M.; Ishikawa,~H.; Seebach,~D.; Hayashi,~Y.
  Organocatalyzed Michael Addition of Aldehydes to Nitro Alkenes - Generally
  Accepted Mechanism Revisited and Revised. \emph{Helv. Chim. Acta}
  \textbf{2011}, \emph{94}, 719--745\relax
\mciteBstWouldAddEndPuncttrue
\mciteSetBstMidEndSepPunct{\mcitedefaultmidpunct}
{\mcitedefaultendpunct}{\mcitedefaultseppunct}\relax
\EndOfBibitem
\bibitem[Bur{\'{e}}s \latin{et~al.}(2016)Bur{\'{e}}s, Armstrong, and
  Blackmond]{Bures2016}
Bur{\'{e}}s,~J.; Armstrong,~A.; Blackmond,~D.~G. Explaining Anomalies in
  Enamine Catalysis: {\textquotedblleft}Downstream Species{\textquotedblright}
  as a New Paradigm for Stereocontrol. \emph{Accounts Chem. Res.}
  \textbf{2016}, \emph{49}, 214--222\relax
\mciteBstWouldAddEndPuncttrue
\mciteSetBstMidEndSepPunct{\mcitedefaultmidpunct}
{\mcitedefaultendpunct}{\mcitedefaultseppunct}\relax
\EndOfBibitem
\bibitem[Seebach \latin{et~al.}(2012)Seebach, Sun, Sparr, Ebert, Schweizer, and
  Beck]{Seebach2012}
Seebach,~D.; Sun,~X.; Sparr,~C.; Ebert,~M.-O.; Schweizer,~W.~B.; Beck,~A.~K.
  1,2-Oxazine N-Oxides as Catalyst Resting States in Michael Additions of
  Aldehydes to Nitro Olefins Organocatalyzed by
  $\alpha$,$\alpha$-Diphenylprolinol Trimethylsilyl Ether. \emph{Helv. Chim.
  Acta} \textbf{2012}, \emph{95}, 1064--1078\relax
\mciteBstWouldAddEndPuncttrue
\mciteSetBstMidEndSepPunct{\mcitedefaultmidpunct}
{\mcitedefaultendpunct}{\mcitedefaultseppunct}\relax
\EndOfBibitem
\bibitem[Gurubrahamam \latin{et~al.}(2016)Gurubrahamam, ming Chen, Huang, Chan,
  Chang, Tsai, and Chen]{Gurubrahamam2016}
Gurubrahamam,~R.; ming Chen,~Y.; Huang,~W.-Y.; Chan,~Y.-T.; Chang,~H.-K.;
  Tsai,~M.-K.; Chen,~K. Dihydrooxazine $N$-Oxide Intermediates as Resting
  States in Organocatalytic Kinetic Resolution of Functionalized Nitroallylic
  Amines with Aldehydes. \emph{Org. Lett.} \textbf{2016}, \emph{18},
  3046--3049\relax
\mciteBstWouldAddEndPuncttrue
\mciteSetBstMidEndSepPunct{\mcitedefaultmidpunct}
{\mcitedefaultendpunct}{\mcitedefaultseppunct}\relax
\EndOfBibitem
\bibitem[Unsleber \latin{et~al.}(2022)Unsleber, Grimmel, and
  Reiher]{Unsleber2022}
Unsleber,~J.~P.; Grimmel,~S.~A.; Reiher,~M. Chemoton 2.0: Autonomous
  Exploration of Chemical Reaction Networks. \emph{J. Chem. Theory Comput.}
  \textbf{2022}, \relax
\mciteBstWouldAddEndPunctfalse
\mciteSetBstMidEndSepPunct{\mcitedefaultmidpunct}
{}{\mcitedefaultseppunct}\relax
\EndOfBibitem
\bibitem[Bensberg \latin{et~al.}(2022)Bensberg, Grimmel, Simm, Sobez, Steiner,
  Türtscher, Unsleber, Weymuth, and Reiher]{Bensberg2022b}
Bensberg,~M.; Grimmel,~S.~A.; Simm,~G.~N.; Sobez,~J.-G.; Steiner,~M.;
  Türtscher,~P.~L.; Unsleber,~J.~P.; Weymuth,~T.; Reiher,~M. qcscine/chemoton:
  Release 2.1.0. 2022; DOI: 10.5281/zenodo.6984579\relax
\mciteBstWouldAddEndPuncttrue
\mciteSetBstMidEndSepPunct{\mcitedefaultmidpunct}
{\mcitedefaultendpunct}{\mcitedefaultseppunct}\relax
\EndOfBibitem
\bibitem[Simm and Reiher(2017)Simm, and Reiher]{Simm2017}
Simm,~G.~N.; Reiher,~M. Context-Driven Exploration of Complex Chemical Reaction
  Networks. \emph{J. Chem. Theory Comput.} \textbf{2017}, \emph{13},
  6108--6119\relax
\mciteBstWouldAddEndPuncttrue
\mciteSetBstMidEndSepPunct{\mcitedefaultmidpunct}
{\mcitedefaultendpunct}{\mcitedefaultseppunct}\relax
\EndOfBibitem
\bibitem[Sobez and Reiher(2020)Sobez, and Reiher]{Sobez2020}
Sobez,~J.-G.; Reiher,~M. \textsc{Molassembler}: Molecular Graph Construction,
  Modification, and Conformer Generation for Inorganic and Organic Molecules.
  \emph{J. Chem. Inf. Model.} \textbf{2020}, \emph{60}, 3884--3900\relax
\mciteBstWouldAddEndPuncttrue
\mciteSetBstMidEndSepPunct{\mcitedefaultmidpunct}
{\mcitedefaultendpunct}{\mcitedefaultseppunct}\relax
\EndOfBibitem
\bibitem[Bannwarth \latin{et~al.}(2019)Bannwarth, Ehlert, and
  Grimme]{Bannwarth2019}
Bannwarth,~C.; Ehlert,~S.; Grimme,~S. GFN2-xTB—An Accurate and Broadly
  Parametrized Self-Consistent Tight-Binding Quantum Chemical Method with
  Multipole Electrostatics and Density-Dependent Dispersion Contributions.
  \emph{J. Chem. Theory Comput.} \textbf{2019}, \emph{15}, 1652--1671\relax
\mciteBstWouldAddEndPuncttrue
\mciteSetBstMidEndSepPunct{\mcitedefaultmidpunct}
{\mcitedefaultendpunct}{\mcitedefaultseppunct}\relax
\EndOfBibitem
\bibitem[Riplinger \latin{et~al.}(2016)Riplinger, Pinski, Becker, Valeev, and
  Neese]{Riplinger2016}
Riplinger,~C.; Pinski,~P.; Becker,~U.; Valeev,~E.~F.; Neese,~F. Sparse
  maps{\textemdash}A systematic infrastructure for reduced-scaling electronic
  structure methods. {II}. Linear scaling domain based pair natural orbital
  coupled cluster theory. \emph{J. Chem. Phys.} \textbf{2016}, \emph{144},
  024109\relax
\mciteBstWouldAddEndPuncttrue
\mciteSetBstMidEndSepPunct{\mcitedefaultmidpunct}
{\mcitedefaultendpunct}{\mcitedefaultseppunct}\relax
\EndOfBibitem
\bibitem[Riplinger \latin{et~al.}(2013)Riplinger, Sandhoefer, Hansen, and
  Neese]{Riplinger2013a}
Riplinger,~C.; Sandhoefer,~B.; Hansen,~A.; Neese,~F. Natural triple excitations
  in local coupled cluster calculations with pair natural orbitals. \emph{J.
  Chem. Phys.} \textbf{2013}, \emph{139}, 134101\relax
\mciteBstWouldAddEndPuncttrue
\mciteSetBstMidEndSepPunct{\mcitedefaultmidpunct}
{\mcitedefaultendpunct}{\mcitedefaultseppunct}\relax
\EndOfBibitem
\bibitem[Reiher(2019)]{Reiher2019}
Reiher,~M. {Mechanistic insight into organic and industrial transformations:
  general discussion}. \emph{Faraday Discuss.} \textbf{2019}, \emph{220},
  299--300\relax
\mciteBstWouldAddEndPuncttrue
\mciteSetBstMidEndSepPunct{\mcitedefaultmidpunct}
{\mcitedefaultendpunct}{\mcitedefaultseppunct}\relax
\EndOfBibitem
\bibitem[Niemeyer and Sung(2014)Niemeyer, and Sung]{Niemeyer2014}
Niemeyer,~K.~E.; Sung,~C.-J. Accelerating moderately stiff chemical kinetics in
  reactive-flow simulations using {GPUs}. \emph{J. Comput. Phys.}
  \textbf{2014}, \emph{256}, 854--871\relax
\mciteBstWouldAddEndPuncttrue
\mciteSetBstMidEndSepPunct{\mcitedefaultmidpunct}
{\mcitedefaultendpunct}{\mcitedefaultseppunct}\relax
\EndOfBibitem
\bibitem[Ismail \latin{et~al.}(2019)Ismail, Stuttaford-Fowler, Ashok,
  Robertson, and Habershon]{Ismail2019}
Ismail,~I.; Stuttaford-Fowler,~H. B. V.~A.; Ashok,~C.~O.; Robertson,~C.;
  Habershon,~S. Automatic Proposal of Multistep Reaction Mechanisms using a
  Graph-Driven Search. \emph{J. Phys. Chem. A} \textbf{2019}, \emph{123},
  3407--3417\relax
\mciteBstWouldAddEndPuncttrue
\mciteSetBstMidEndSepPunct{\mcitedefaultmidpunct}
{\mcitedefaultendpunct}{\mcitedefaultseppunct}\relax
\EndOfBibitem
\bibitem[Rappoport \latin{et~al.}(2014)Rappoport, Galvin, Zubarev, and
  Aspuru-Guzik]{Rappoport2014}
Rappoport,~D.; Galvin,~C.~J.; Zubarev,~D.~Y.; Aspuru-Guzik,~A. Complex Chemical
  Reaction Networks from Heuristics-Aided Quantum Chemistry. \emph{J. Chem.
  Theory Comput.} \textbf{2014}, \emph{10}, 897--907\relax
\mciteBstWouldAddEndPuncttrue
\mciteSetBstMidEndSepPunct{\mcitedefaultmidpunct}
{\mcitedefaultendpunct}{\mcitedefaultseppunct}\relax
\EndOfBibitem
\bibitem[Bergeler \latin{et~al.}(2015)Bergeler, Simm, Proppe, and
  Reiher]{Bergeler2015}
Bergeler,~M.; Simm,~G.~N.; Proppe,~J.; Reiher,~M. Heuristics-Guided Exploration
  of Reaction Mechanisms. \emph{J. Chem. Theory Comput.} \textbf{2015},
  \emph{11}, 5712--5722\relax
\mciteBstWouldAddEndPuncttrue
\mciteSetBstMidEndSepPunct{\mcitedefaultmidpunct}
{\mcitedefaultendpunct}{\mcitedefaultseppunct}\relax
\EndOfBibitem
\bibitem[Grimmel and Reiher(2019)Grimmel, and Reiher]{Grimmel2019}
Grimmel,~S.~A.; Reiher,~M. The electrostatic potential as a descriptor for the
  protonation propensity in automated exploration of reaction mechanisms.
  \emph{Faraday Discuss.} \textbf{2019}, \emph{220}, 443--463\relax
\mciteBstWouldAddEndPuncttrue
\mciteSetBstMidEndSepPunct{\mcitedefaultmidpunct}
{\mcitedefaultendpunct}{\mcitedefaultseppunct}\relax
\EndOfBibitem
\bibitem[Grimmel and Reiher(2021)Grimmel, and Reiher]{Grimmel2021}
Grimmel,~S.~A.; Reiher,~M. On the Predictive Power of Chemical Concepts.
  \emph{{CHIMIA}} \textbf{2021}, \emph{75}, 311\relax
\mciteBstWouldAddEndPuncttrue
\mciteSetBstMidEndSepPunct{\mcitedefaultmidpunct}
{\mcitedefaultendpunct}{\mcitedefaultseppunct}\relax
\EndOfBibitem
\bibitem[Vaucher and Reiher(2018)Vaucher, and Reiher]{Vaucher2018}
Vaucher,~A.~C.; Reiher,~M. Minimum Energy Paths and Transition States by Curve
  Optimization. \emph{J. Chem. Theory Comput.} \textbf{2018}, \emph{14},
  3091--3099\relax
\mciteBstWouldAddEndPuncttrue
\mciteSetBstMidEndSepPunct{\mcitedefaultmidpunct}
{\mcitedefaultendpunct}{\mcitedefaultseppunct}\relax
\EndOfBibitem
\bibitem[Bensberg \latin{et~al.}(2022)Bensberg, Brunken, Csizi, Grimmel,
  Gugler, Sobez, Steiner, Türtscher, Unsleber, Weymuth, and
  Reiher]{Bensberg2022d}
Bensberg,~M.; Brunken,~C.; Csizi,~K.-S.; Grimmel,~S.~A.; Gugler,~S.;
  Sobez,~J.-G.; Steiner,~M.; Türtscher,~P.~L.; Unsleber,~J.~P.; Weymuth,~T.;
  Reiher,~M. qcscine/puffin: Release 1.1.0. 2022; DOI:
  10.5281/zenodo.6984580\relax
\mciteBstWouldAddEndPuncttrue
\mciteSetBstMidEndSepPunct{\mcitedefaultmidpunct}
{\mcitedefaultendpunct}{\mcitedefaultseppunct}\relax
\EndOfBibitem
\bibitem[Brunken \latin{et~al.}(2022)Brunken, Csizi, Grimmel, Gugler, Sobez,
  Steiner, Türtscher, Unsleber, Vaucher, Weymuth, and Reiher]{Brunken2022}
Brunken,~C.; Csizi,~K.-S.; Grimmel,~S.~A.; Gugler,~S.; Sobez,~J.-G.;
  Steiner,~M.; Türtscher,~P.~L.; Unsleber,~J.~P.; Vaucher,~A.~C.; Weymuth,~T.;
  Reiher,~M. qcscine/readuct: Release 4.1.0. 2022; DOI:
  10.5281/zenodo.6984575\relax
\mciteBstWouldAddEndPuncttrue
\mciteSetBstMidEndSepPunct{\mcitedefaultmidpunct}
{\mcitedefaultendpunct}{\mcitedefaultseppunct}\relax
\EndOfBibitem
\bibitem[Onufriev \latin{et~al.}(2004)Onufriev, Bashford, and
  Case]{Onufriev2004}
Onufriev,~A.; Bashford,~D.; Case,~D.~A. Exploring protein native states and
  large-scale conformational changes with a modified generalized born model.
  \emph{Proteins: Struct., Funct., Bioinf.} \textbf{2004}, \emph{55},
  383--394\relax
\mciteBstWouldAddEndPuncttrue
\mciteSetBstMidEndSepPunct{\mcitedefaultmidpunct}
{\mcitedefaultendpunct}{\mcitedefaultseppunct}\relax
\EndOfBibitem
\bibitem[Sigalov \latin{et~al.}(2006)Sigalov, Fenley, and
  Onufriev]{Sigalov2006}
Sigalov,~G.; Fenley,~A.; Onufriev,~A. Analytical electrostatics for
  biomolecules: Beyond the generalized Born approximation. \emph{J. Chem.
  Phys.} \textbf{2006}, \emph{124}, 124902\relax
\mciteBstWouldAddEndPuncttrue
\mciteSetBstMidEndSepPunct{\mcitedefaultmidpunct}
{\mcitedefaultendpunct}{\mcitedefaultseppunct}\relax
\EndOfBibitem
\bibitem[Ahlrichs \latin{et~al.}(1989)Ahlrichs, B{\"a}r, H{\"a}ser, Horn, and
  K{\"o}lmel]{Ahlrichs1989}
Ahlrichs,~R.; B{\"a}r,~M.; H{\"a}ser,~M.; Horn,~H.; K{\"o}lmel,~C. Electronic
  structure calculations on workstation computers: The program system
  turbomole. \emph{Chem. Phys. Lett.} \textbf{1989}, \emph{162}, 165--169\relax
\mciteBstWouldAddEndPuncttrue
\mciteSetBstMidEndSepPunct{\mcitedefaultmidpunct}
{\mcitedefaultendpunct}{\mcitedefaultseppunct}\relax
\EndOfBibitem
\bibitem[tur()]{turbomole741}
{TURBOMOLE V7.4.2 2019}, a development of {University of Karlsruhe} and
  {Forschungszentrum Karlsruhe GmbH}, 1989-2007, {TURBOMOLE GmbH}, since 2007;
  available from {\tt http://www.turbomole.com}.\relax
\mciteBstWouldAddEndPunctfalse
\mciteSetBstMidEndSepPunct{\mcitedefaultmidpunct}
{}{\mcitedefaultseppunct}\relax
\EndOfBibitem
\bibitem[Perdew \latin{et~al.}(1996)Perdew, Burke, and Ernzerhof]{Perdew96}
Perdew,~J.~P.; Burke,~K.; Ernzerhof,~M. {Generalized Gradient Approximation
  Made Simple}. \emph{Phys. Rev. Lett.} \textbf{1996}, \emph{77}, 3865\relax
\mciteBstWouldAddEndPuncttrue
\mciteSetBstMidEndSepPunct{\mcitedefaultmidpunct}
{\mcitedefaultendpunct}{\mcitedefaultseppunct}\relax
\EndOfBibitem
\bibitem[Grimme \latin{et~al.}(2010)Grimme, Antony, Ehrlich, and
  Krieg]{Grimme2010}
Grimme,~S.; Antony,~J.; Ehrlich,~S.; Krieg,~H. {A consistent and accurate ab
  initio parametrization of density functional dispersion correction (DFT-D)
  for the 94 elements H-Pu}. \emph{J. Chem. Phys.} \textbf{2010}, \emph{132},
  154104\relax
\mciteBstWouldAddEndPuncttrue
\mciteSetBstMidEndSepPunct{\mcitedefaultmidpunct}
{\mcitedefaultendpunct}{\mcitedefaultseppunct}\relax
\EndOfBibitem
\bibitem[Grimme \latin{et~al.}(2011)Grimme, Ehrlich, and Goerigk]{Grimme2011}
Grimme,~S.; Ehrlich,~S.; Goerigk,~L. {Effect of the damping function in
  dispersion corrected density functional theory}. \emph{J. Comput. Chem.}
  \textbf{2011}, \emph{32}, 1456--1465\relax
\mciteBstWouldAddEndPuncttrue
\mciteSetBstMidEndSepPunct{\mcitedefaultmidpunct}
{\mcitedefaultendpunct}{\mcitedefaultseppunct}\relax
\EndOfBibitem
\bibitem[Weigend and Ahlrichs(2005)Weigend, and Ahlrichs]{Ahlrich2005}
Weigend,~F.; Ahlrichs,~R. {Balanced basis sets of split valence, triple zeta
  valence and quadruple zeta valence quality for H to Rn: Design and assessment
  of accuracy}. \emph{Phys. Chem. Chem. Phys.} \textbf{2005}, \emph{7},
  3297--3305\relax
\mciteBstWouldAddEndPuncttrue
\mciteSetBstMidEndSepPunct{\mcitedefaultmidpunct}
{\mcitedefaultendpunct}{\mcitedefaultseppunct}\relax
\EndOfBibitem
\bibitem[Klamt and Schüürmann(1993)Klamt, and Schüürmann]{Klamt1993}
Klamt,~A.; Schüürmann,~G. {COSMO}: a new approach to dielectric screening in
  solvents with explicit expressions for the screening energy and its gradient.
  \emph{J Chem. Soc., Perkin Trans 2} \textbf{1993}, 799--805\relax
\mciteBstWouldAddEndPuncttrue
\mciteSetBstMidEndSepPunct{\mcitedefaultmidpunct}
{\mcitedefaultendpunct}{\mcitedefaultseppunct}\relax
\EndOfBibitem
\bibitem[Klamt(2011)]{Klamt2011}
Klamt,~A. The {COSMO} and {COSMO}-{RS} solvation models. \emph{WIREs Comput Mol
  Sci} \textbf{2011}, \emph{1}, 699--709\relax
\mciteBstWouldAddEndPuncttrue
\mciteSetBstMidEndSepPunct{\mcitedefaultmidpunct}
{\mcitedefaultendpunct}{\mcitedefaultseppunct}\relax
\EndOfBibitem
\bibitem[Proppe and Reiher(2022)Proppe, and Reiher]{Proppe2022}
Proppe,~J.; Reiher,~M. qcscine/kinetx: Release 1.0.0. 2022; DOI:
  10.5281/ZENODO.6471956\relax
\mciteBstWouldAddEndPuncttrue
\mciteSetBstMidEndSepPunct{\mcitedefaultmidpunct}
{\mcitedefaultendpunct}{\mcitedefaultseppunct}\relax
\EndOfBibitem
\bibitem[Pihko \latin{et~al.}(2004)Pihko, Nyberg, and Usano]{Pihko2004}
Pihko,~P.; Nyberg,~A.; Usano,~A. {Proline-Catalyzed Ketone-Aldehyde Aldol
  Reactions are Accelerated by Water}. \emph{Synlett} \textbf{2004},
  \emph{2004}, 1891--1896\relax
\mciteBstWouldAddEndPuncttrue
\mciteSetBstMidEndSepPunct{\mcitedefaultmidpunct}
{\mcitedefaultendpunct}{\mcitedefaultseppunct}\relax
\EndOfBibitem
\bibitem[Pihko \latin{et~al.}(2006)Pihko, Laurikainen, Usano, Nyberg, and
  Kaavi]{Pihko2006}
Pihko,~P.~M.; Laurikainen,~K.~M.; Usano,~A.; Nyberg,~A.~I.; Kaavi,~J.~A.
  {Effect of additives on the proline-catalyzed ketone{\textendash}aldehyde
  aldol reactions}. \emph{Tetrahedron} \textbf{2006}, \emph{62}, 317--328\relax
\mciteBstWouldAddEndPuncttrue
\mciteSetBstMidEndSepPunct{\mcitedefaultmidpunct}
{\mcitedefaultendpunct}{\mcitedefaultseppunct}\relax
\EndOfBibitem
\bibitem[Zotova \latin{et~al.}(2007)Zotova, Franzke, Armstrong, and
  Blackmond]{Zotova2007}
Zotova,~N.; Franzke,~A.; Armstrong,~A.; Blackmond,~D.~G. {Clarification of the
  Role of Water in Proline-Mediated Aldol Reactions}. \emph{J. Am. Chem. Soc.}
  \textbf{2007}, \emph{129}, 15100--15101\relax
\mciteBstWouldAddEndPuncttrue
\mciteSetBstMidEndSepPunct{\mcitedefaultmidpunct}
{\mcitedefaultendpunct}{\mcitedefaultseppunct}\relax
\EndOfBibitem
\bibitem[Bensberg \latin{et~al.}(2022)Bensberg, Brandino, Can, Del, Grimmel,
  Mesiti, Müller, Steiner, Türtscher, Unsleber, Weberndorfer, Weymuth, and
  Reiher]{Bensberg2022c}
Bensberg,~M.; Brandino,~G.~P.; Can,~Y.; Del,~M.; Grimmel,~S.~A.; Mesiti,~M.;
  Müller,~C.~H.; Steiner,~M.; Türtscher,~P.~L.; Unsleber,~J.~P.;
  Weberndorfer,~M.; Weymuth,~T.; Reiher,~M. qcscine/heron: Release 1.0.0. 2022;
  DOI: 10.5281/zenodo.7038388\relax
\mciteBstWouldAddEndPuncttrue
\mciteSetBstMidEndSepPunct{\mcitedefaultmidpunct}
{\mcitedefaultendpunct}{\mcitedefaultseppunct}\relax
\EndOfBibitem
\bibitem[Barone and Cossi(1998)Barone, and Cossi]{Barone1998}
Barone,~V.; Cossi,~M. Quantum Calculation of Molecular Energies and Energy
  Gradients in Solution by a Conductor Solvent Model. \emph{J. Phys. Chem. A}
  \textbf{1998}, \emph{102}, 1995--2001\relax
\mciteBstWouldAddEndPuncttrue
\mciteSetBstMidEndSepPunct{\mcitedefaultmidpunct}
{\mcitedefaultendpunct}{\mcitedefaultseppunct}\relax
\EndOfBibitem
\bibitem[Neese(2017)]{Neese2017}
Neese,~F. Software update: the {ORCA} program system, version 4.0. \emph{WIREs
  Comput Mol Sci} \textbf{2017}, \emph{8}, e1327\relax
\mciteBstWouldAddEndPuncttrue
\mciteSetBstMidEndSepPunct{\mcitedefaultmidpunct}
{\mcitedefaultendpunct}{\mcitedefaultseppunct}\relax
\EndOfBibitem
\bibitem[Uhe \latin{et~al.}(2010)Uhe, Kozuch, and Shaik]{Uhe2010}
Uhe,~A.; Kozuch,~S.; Shaik,~S. Automatic analysis of computed catalytic cycles.
  \emph{J Comput. Chem.} \textbf{2010}, \emph{32}, 978--985\relax
\mciteBstWouldAddEndPuncttrue
\mciteSetBstMidEndSepPunct{\mcitedefaultmidpunct}
{\mcitedefaultendpunct}{\mcitedefaultseppunct}\relax
\EndOfBibitem
\bibitem[Unsleber(2023)]{Unsleber2023}
Unsleber,~J.~P. {Accelerating Reaction Network Explorations with Automated
  Reaction Template Extraction and Application}. \emph{ChemRxiv} \textbf{2023},
  DOI: 10.26434/chemrxiv-2023-lgnrm\relax
\mciteBstWouldAddEndPuncttrue
\mciteSetBstMidEndSepPunct{\mcitedefaultmidpunct}
{\mcitedefaultendpunct}{\mcitedefaultseppunct}\relax
\EndOfBibitem
\bibitem[Bensberg and Reiher(2023)]{Bensberg2023c}
Bensberg,~M.; Reiher,~M. PROLINE Exploration Data Set; DOI: 10.5281/7703748\relax
\mciteBstWouldAddEndPuncttrue
\mciteSetBstMidEndSepPunct{\mcitedefaultmidpunct}
{\mcitedefaultendpunct}{\mcitedefaultseppunct}\relax
\EndOfBibitem
\end{mcitethebibliography}
\end{document}